\documentclass[aps,pre,twocolumn,showpacs,superscriptaddress,groupedaddress,nofootinbib]{revtex4-1}  % for review and submission

\usepackage{graphicx}

\newcommand{\be}{ \begin{equation} }
\newcommand{\ee}{ \end{equation} }
\newcommand{\bea}{\begin{eqnarray} \nonumber }
\newcommand{\eea}{\end{eqnarray}}
\newcommand{\bi}{\begin{itemize}}
\newcommand{\ei}{\end{itemize}}

\newcommand{\cRM}[1]{\MakeUppercase{\romannumeral #1}} % Capital
\begin{document}

%\preprint{}

\title{Universal size effects for populations in group-outcome decision-making problems}


\author{Christian Borghesi$^{1}$, Laura Hern\'andez$^{1}$, R\'emi Louf$^{2}$, Fabrice Caparros$^{3}$}
\email{christian.borghesi@u-cergy.fr}
\affiliation{\small$^{1}$ Laboratoire de Physique Th\'eorique et Mod\'elisation, UMR-8089 CNRS-Universit\'e Cergy Pontoise, France\\
$^{2}$ Institut de Physique Th\'eorique, CEA-Saclay, France\\
$^{3}$ D\'epartement G\'eographie, Centre Universitaire de Formation et de Recherche de Mayotte, France}
\date{\today}

\begin{abstract}
Elections constitute a paradigm of decision-making problems that have puzzled experts of different disciplines for decades. We study two decision-making problems, where groups make decisions that impact only themselves as a group. In both studied cases, participation in local elections and the number of democratic representatives at different scales (from local to national), we observe a universal scaling with the constituency size. These results may be interpreted as constituencies having a hierarchical structure, where each group of $N$ agents, at each level of the hierarchy, is divided in about $N^{\delta}$ subgroups with $\delta \approx 1/3$. Following this interpretation, we propose a phenomenological model of vote participation where abstention is related to the perceived link of an agent to the rest of the constituency and which reproduces quantitatively the observed data.
\end{abstract}

% 'we propose': direct tense is better


%\keywords{}
%\bf Keywords}: universal scaling $|$ democratic decision-making $|$ group-size effects $|$ empirical study of turnout.\\

\maketitle

\section{Introduction}

Elections are an excellent tool with which to study decision-making processes. Unlike declarative opinion polls where the obtained data may be affected by different biases, elections are an occasion where objective and reliable data about the opinion of a significant number of people are gathered. If we were to see the opinion dynamics as a natural process, the results of elections would play the role of observed quantities.

Traditionally, studies of opinion dynamics in the physics community are posed as a {\it direct} problem. Authors make assumptions on the way people interact with each other and consequently build a model of opinion spreading based on these assumptions~\cite{fortunato,Bouchaud-rev}. These models successfully describe the qualitative evolution of opinion at a macroscopic level.

More recently, the community has undertaken to study {\it inverse} problems in opinion dynamics. Several authors~\cite{costa_filho_scaling_vot,lyra_bresil_el,fortunato2,daisy-model,araripe_role_parties,diffusive1,araujo_tactical_voting,universality_candidates,diffusive2,weak-law,fortunato-2012,universality_candidates2} have focused on data in the hope of finding regularities, often called stylized facts, before trying to find the underlying mechanisms leading to them. For example, it has been found that the distribution of the number of votes received by candidates is a universal scaling function, suggesting that the global opinion forms in a branching process~\cite{fortunato2}. Also, spatial and temporal regularities observed in the results of French elections~\cite{diffusive1} and the strong spatial correlations observed in the turnout rate of elections in many different countries~\cite{diffusive2}, may be interpreted in terms of the existence of a `cultural field' whose dynamics obeys a two-dimensional random diffusion equation. So far, the observed regularities indicate that the processes at stake in opinion formation are similar, irrespective of the social, cultural and economical background of the voters.

In this work we present results regarding two different decision-making processes. First we consider the participation to local elections (elections of the Mayor or municipal councillors). We have gathered and analyzed data from 21 elections in 10 different countries (see Tab.~\ref{tdata}) with different social, economical, and cultural characteristics. The regularities observed in these data lead us to study the number of representatives to different chambers in different countries (from local to national level). Finally, we propose a phenomenological model which accounts for the empirical regularities and brings a new outlook on the reasons behind abstention in local elections.

%%%%%%%%%%%%%%%%%%%%%%%%%%%%%%%%%%%
\section{Analysis of local elections}

By local elections, we mean elections in municipal districts which go from small villages of hundreds of inhabitants to large cities of hundreds of thousand inhabitants. (Details on data analysis are given in the Appendix; data may be downloaded from Ref.~\cite{download}).

The elections chosen for this study satisfy the following properties: (1) voting is not compulsory, (2) there are enough available data so as to ensure good statistics (typically more than 2000 municipalities), (3) the elections concern only local issues, (4) the number of registered voters is well known.

For each municipality and each election, we denote by $N$ the total number of registered voters, $N_{+}$ the total turnout and $N_{-}$ the total number of abstentions ($N = N_{+} + N_{-} $). We define the {\it logarithmic turnout rate} (LTR) as $\tau = \ln \left( \frac{N_+}{N_-} \right)$, a symmetric and unbounded variable, that highlights statistical regularities appearing over different elections and countries~\cite{diffusive1,diffusive2,these,klimek-irregularities}. 

%Tab. 1
\begin{table}%[tbp]
\caption{\small Local elections analyzed in this paper. The \textit{Mun} column indicates the number of municipalities involved in the election (rounded up or down to the nearest 100, except for Israel); the \textit{Kind} column indicates whether there is only a single round (m) or two, in which case m1 and m2 mean first or the second round respectively. Finally the \textit{Years} column gives the year of the studied elections for each country. In the particular case of Israel, (J) indicates municipalities where the majority of population is Jewish and (A) those where  most of the population is Arab. The horizontal line separate elections that show a deviation from the universal behavior (see text).}
\begin{tabular}{llll}
\hline
\hline
Country & Mun & Kind & Years\\ 
\hline
Austria (At) & 2400 & m & 2004, 2009 \\
Costa Rica (CR) & 1900 & m & 2002, 2006\\
Czech Rep. (Cz) & 6400 & m & 2002, 2006, 2010\\
France (Fr) & 36000 & m1 & 2001, 2008\\
Poland (Pl) & 750 & m2 & 2006, 2010\\
Romania (Ro) & 3200 & m1 & 2012\\
Slovakia (Sk) & 2900 & m & 2006, 2010\\
Spain (Sp) & 2000 & m & 1991, 1995, 1999\\
 & & & 2003, 2007, 2011\\
Israel (Is) (J) & 100 & m & 2008\\
\hline
Portugal (Pt) & 4100 & m & 2009\\
Spain (Sp) & 2000 & m & 1987 \\ 
Israel (Is) (A) & 50 & m & 2008\\ 
\hline
\hline
\end{tabular}
\label{tdata}
\end{table}

%Fig. 1
\begin{figure}
\includegraphics[width=8cm, height=6.5cm,clip=true]{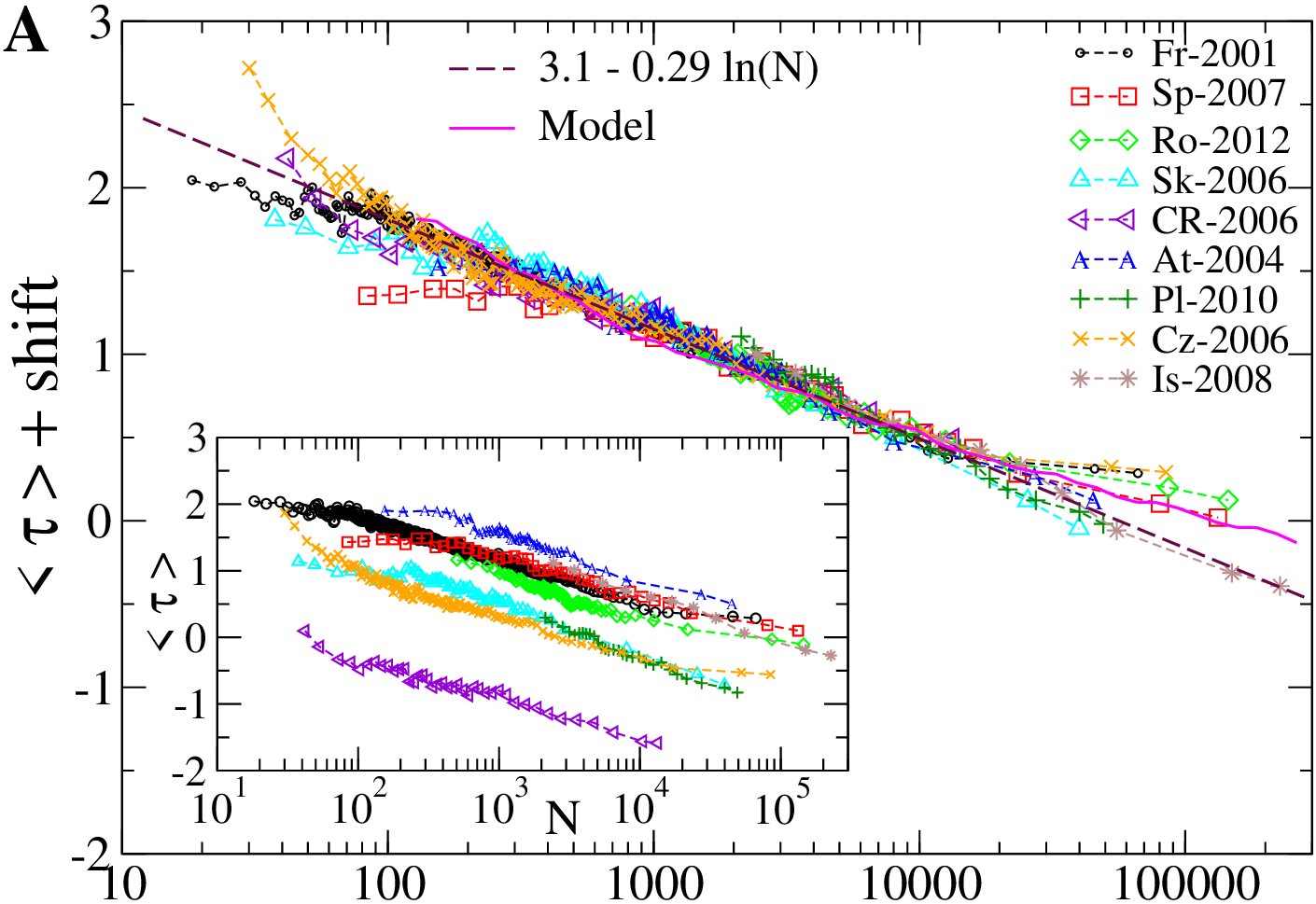}\\
\includegraphics[width=8cm, height=5.5cm,clip=true]{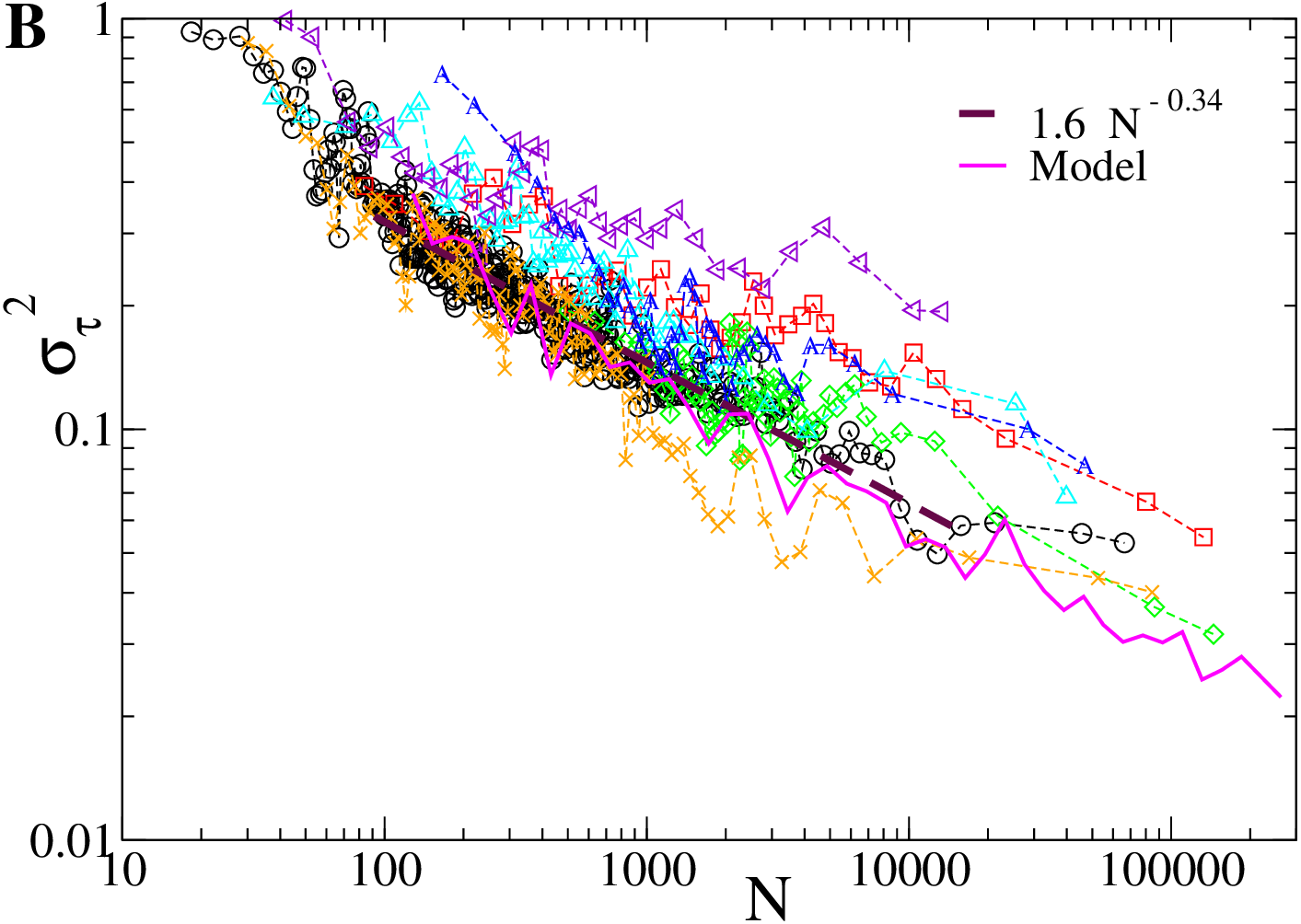} 
\caption{\small (Color online) LTR, $\tau$ as a function of the municipality size $N$ for different local elections (for the sake of clarity we plot only one election for each country; the remaining ones may be found in Fig.~\ref{si_tau}). An election is identified by the country and the corresponding year. ({\bf A}) Average value, $\langle \tau \rangle (N)$. In order to better show the universality of the scaling as a function of $N$, curves have been shifted to make them coincident with the French election (the original curves are shown in the inset). In Israel, only municipalities with Jewish majority are shown. ({\bf B}) variance, $\sigma^2_\tau (N)$ of the conditional distribution $P(\tau|N)$. Due to poor statistics, $\sigma^2_\tau (N)$ is shown for neither Israeli nor Polish elections. See the Appendix for more details on the data analysis. }
\label{ftau}
\end{figure}

We first study the dependence of the average LTR, the first moment of $P\left(\tau|N\right)$, with the size of the considered municipality (Fig.~\ref{ftau}A). Each point corresponds to the average value of $\tau$ over municipalities of about the same size  (see the Appendix). 
The behavior of the average LTR of each election can be correctly fitted by:

\begin{equation} 
\langle \tau \rangle (N) \approx C - \,\alpha\, \ln(N)\,,
\label{emoy}
\end{equation}
where $C$ is an election-dependent constant. Strikingly, the shape of $\langle \tau \rangle (N)$ does not depend on the country that is being studied. Moreover, we find that  the slope $\alpha$ is the same for most of the studied elections, fluctuating around an average value of  $\alpha \approx 0.31$, as is shown in Table~\ref{tstat}. This behavior holds for elections that have a global turnout rate ranging from $18\%$ to $88\%$, implying that the observed behavior is independent of the total participation to a given election. The quality of the fit, excluding very small and very large sizes,  is very good ($R^2 >0.95$).

Interestingly, the variance  $\sigma^2_{\tau}(N)$ of $P\left( \tau | N \right)$ also exhibits regularities, even though the curves are noisier than for the first moment (Fig.~\ref{ftau}B). We fit these curves assuming a power-law dependency with constituency size and we find
\begin{equation}
\sigma^2_\tau (N) \approx \frac{D}{N^\beta}\,, \label{var}
\end{equation}
where $D$ is a constant of the order of unity which depends on the election. The exponent, $\beta$, is the same for most of studied elections and fluctuates around an average value of  $\beta \approx 0.32$ (see Tab.~\ref{tstat}).

%Fig. 2
\begin{figure*}
\includegraphics[width=4.75cm, height=3.7cm,clip=true]{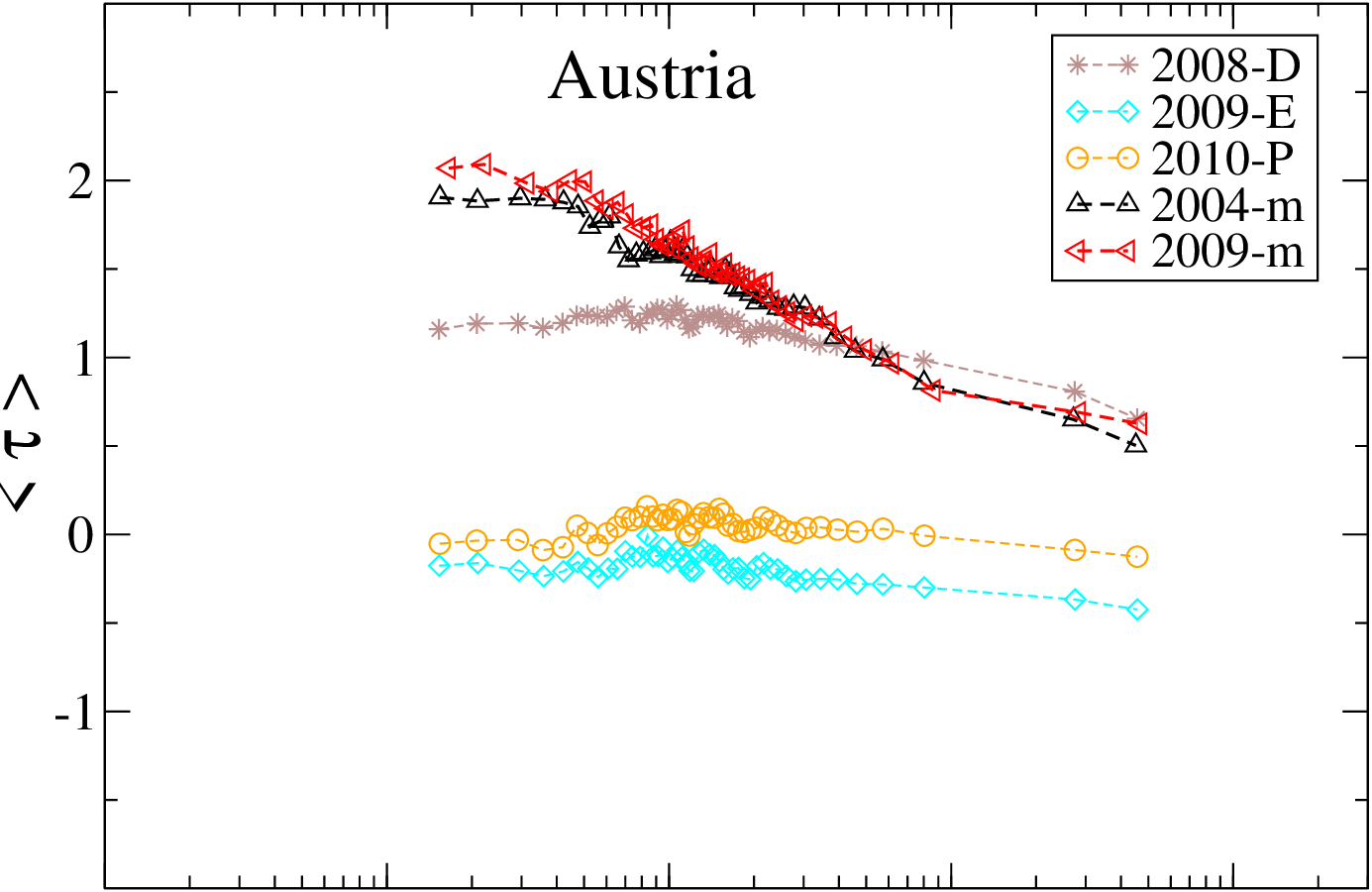}%{si-at.eps}[width=4.25cm, height=4cm,clip=true]
 \includegraphics[width=4.2cm, height=3.7cm,clip=true]{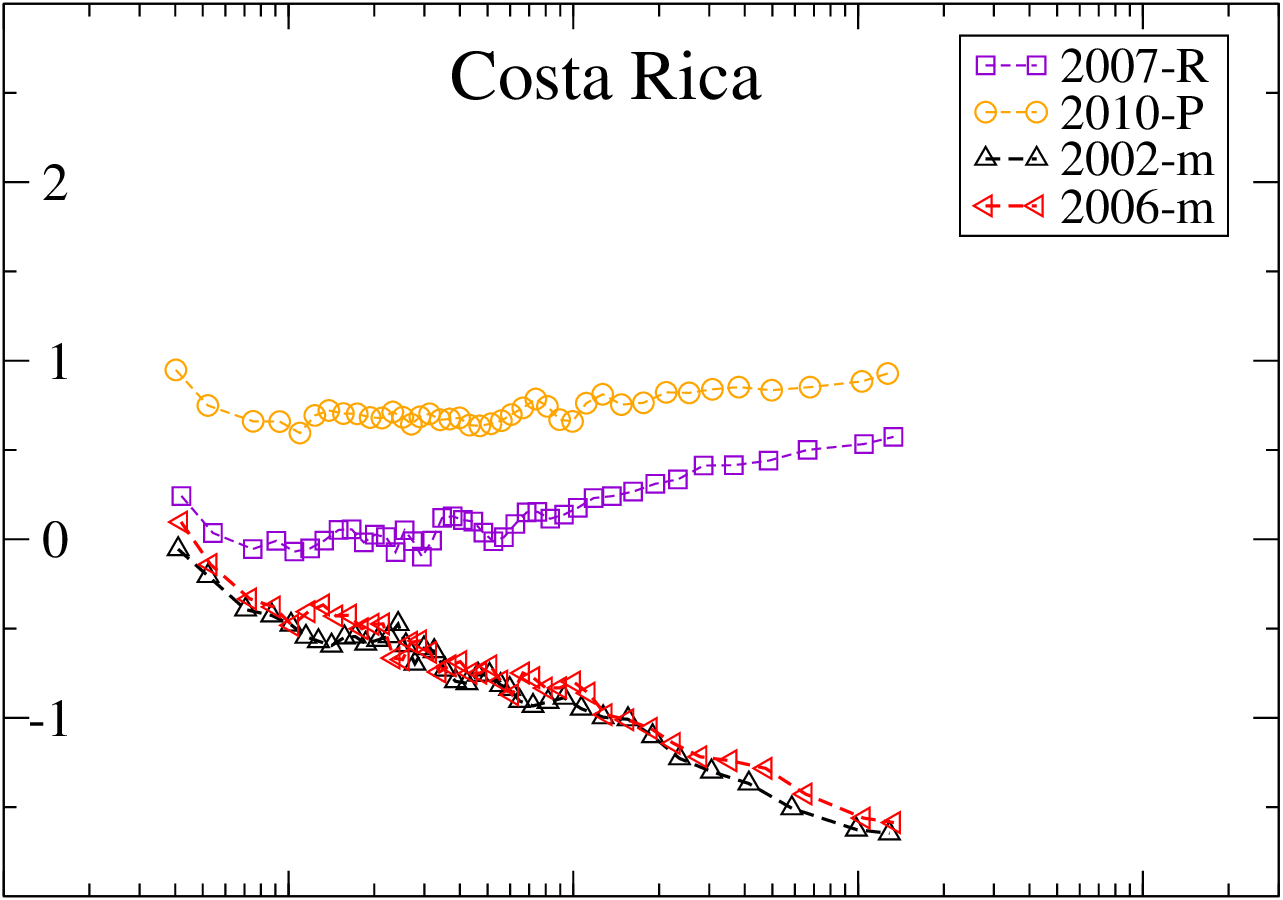}%{si-cr.eps} 
 \includegraphics[width=4.2cm, height=3.7cm,clip=true]{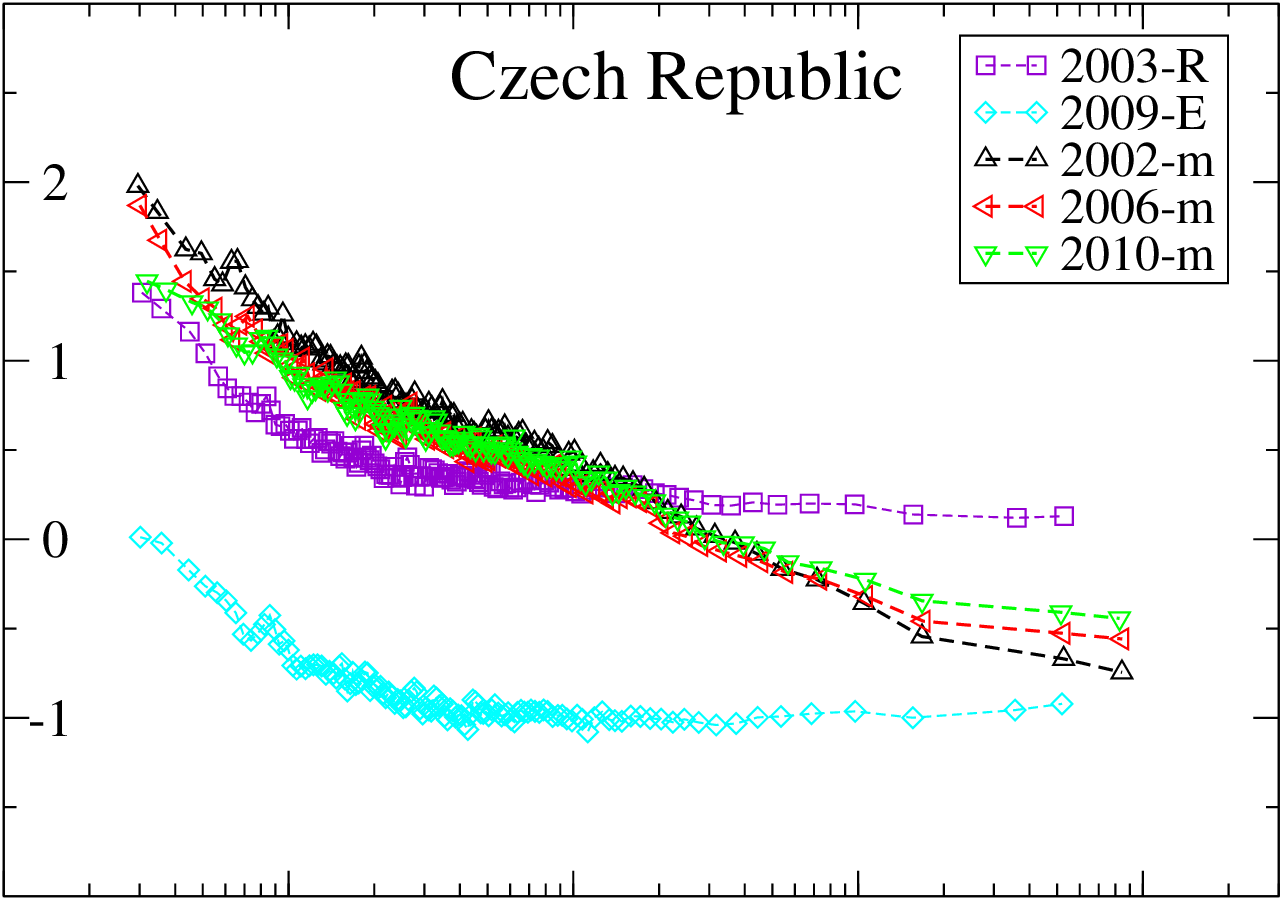}%{si-cz.eps} \\
\includegraphics[width=4.2cm, height=3.7cm,clip=true]{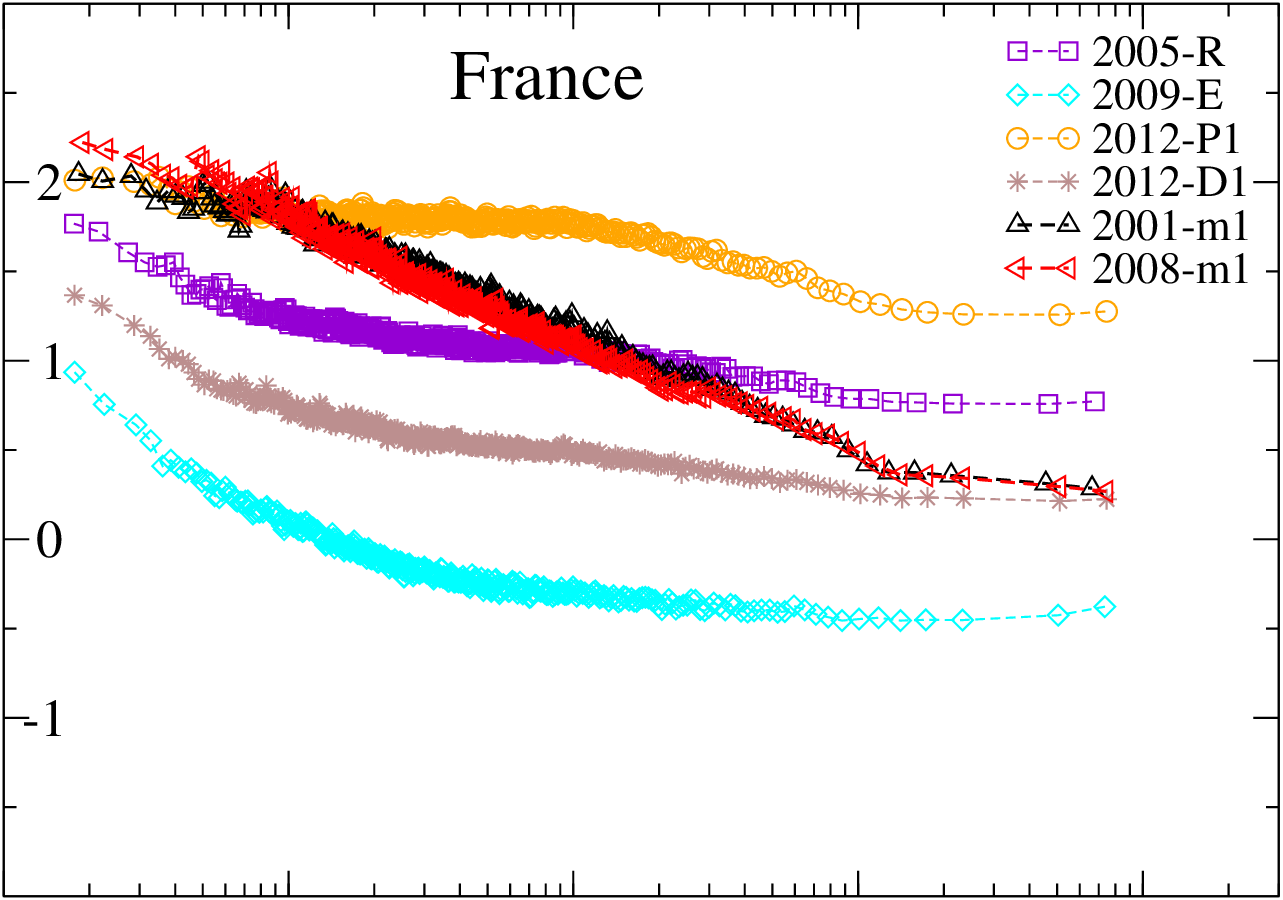}\\%{si-fr.eps} 
 \includegraphics[width=4.75cm, height=4.25cm,clip=true]{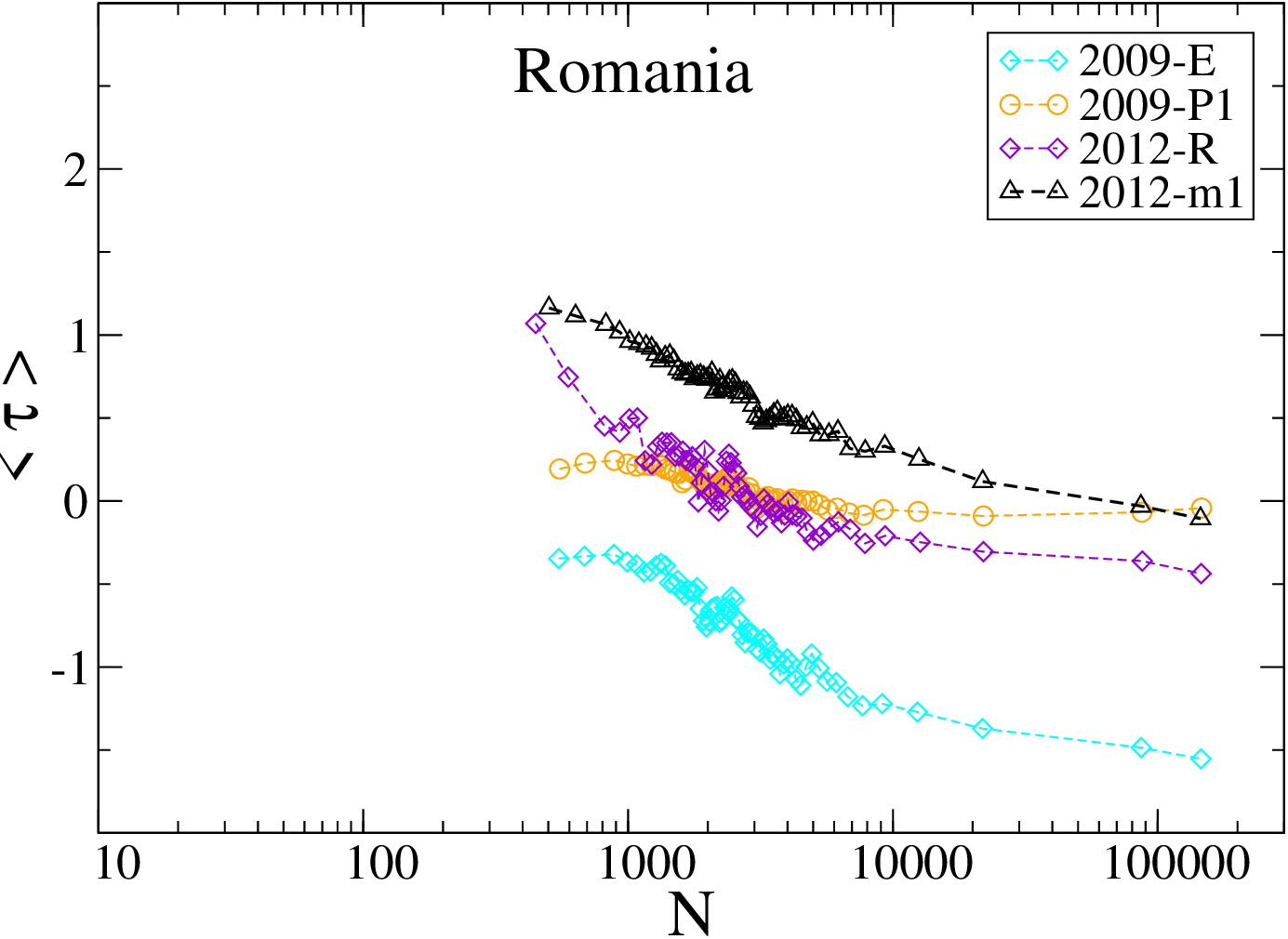}%{si-ro.eps}
 \includegraphics[width=4.2cm, height=4.25cm,clip=true]{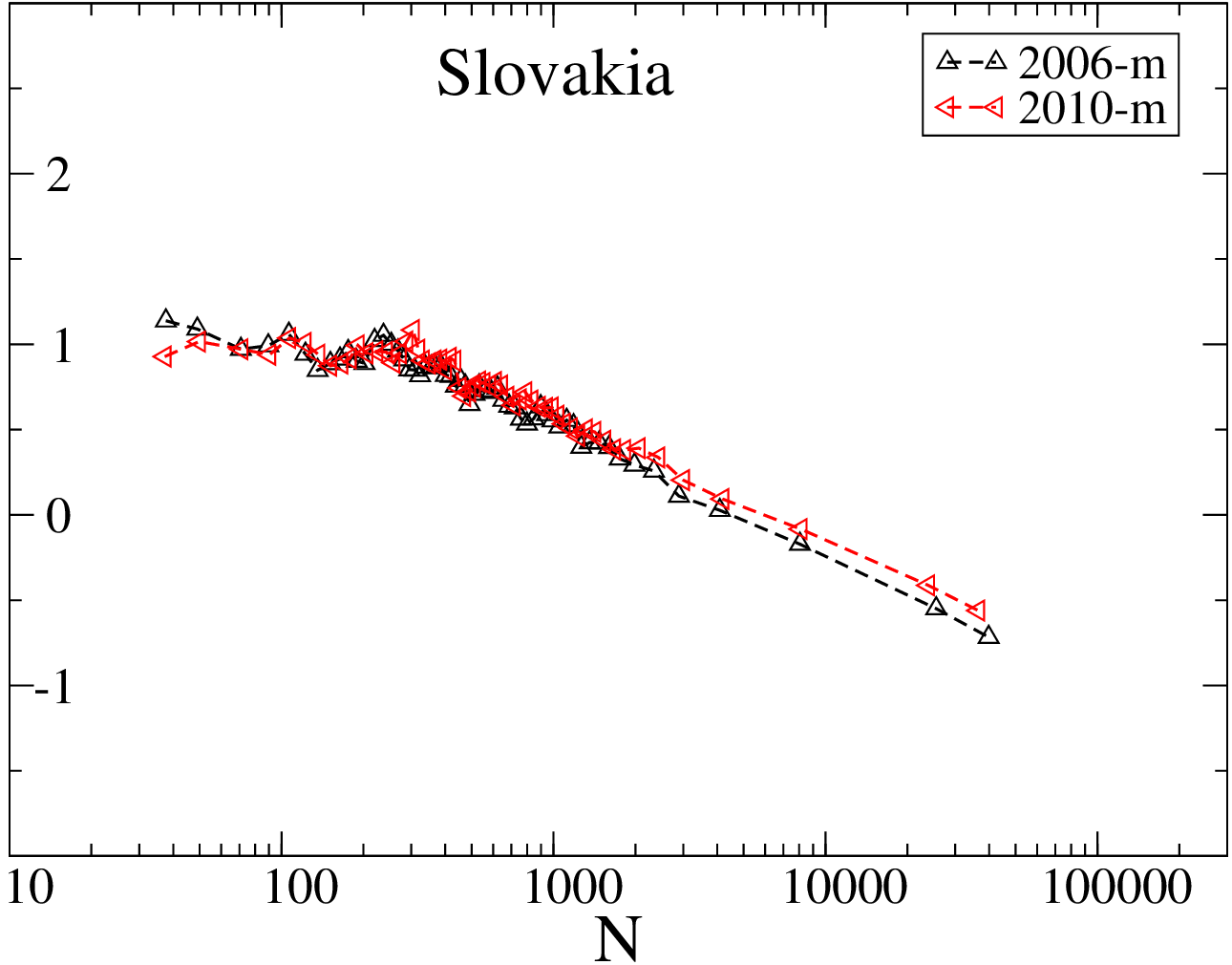}%{si-sk.eps} \\
\includegraphics[width=4.2cm, height=4.25cm,clip=true]{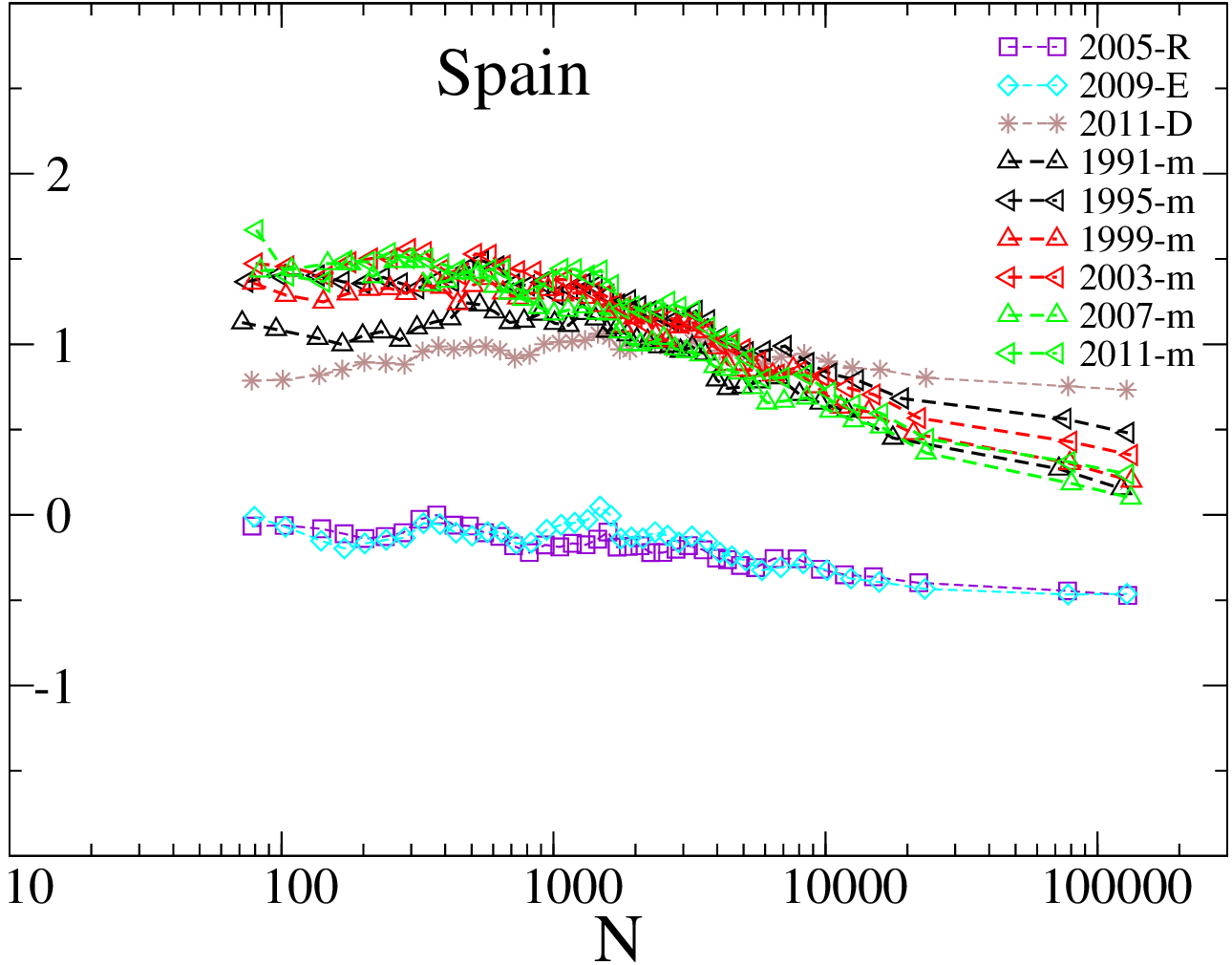}%{si-sp.eps}
 \includegraphics[width=4.2cm, height=4.25cm,clip=true]{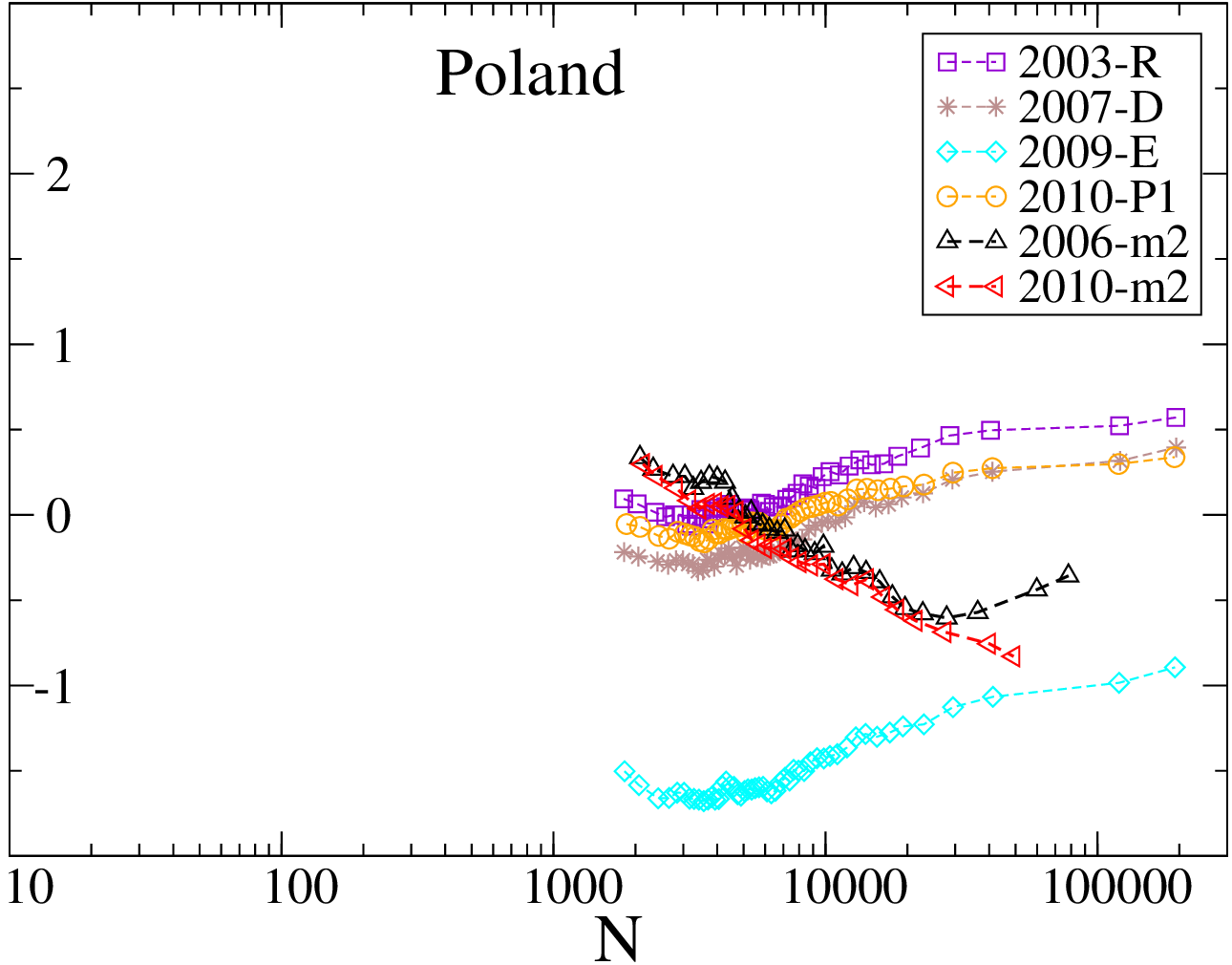}%{si-pl.eps}
\caption{\small (Color online) Comparison of the behavior of $\langle \tau \rangle$ for local and national elections, as a function of the municipality size, $N$. An election is identified by the country, the corresponding year and its type m: Municipal election (in black, red and green), E: European parliament election (in cyan), P: Presidential election (in orange), D: Parlement election (in brown), R: referendum (in violet). Averages have been calculated over the same number of municipalities as in Fig.~\ref{ftau}, with the exception of national Polish elections, where the statistics is better than for local elections.} \label{si_tau}
\end{figure*}

As expected, the quality of the fit of $\sigma^2_{\tau}(N)$ is worse than that of $\langle \tau \rangle (N)$. Moreover, each point of the curve of Fig.~\ref{ftau}B corresponds to the variance of LTR over municipalities of a given size. Because this number is characteristic of each country, the quality of the statistics cannot be  improved. Without surprise, the best fit corresponds to the case of France where each point corresponds to 200 municipalities of very similar size and where nevertheless, the number of obtained points is large (see the Appendix for details on all elections). We have compared this fit with other plausible decreasing functions  like the exponential, the stretched exponential, and the inverse function. For all these functions of two parameters, the obtained fit was clearly worse. We have also tried a logarithmic binning, which does not change the nature of the results.

%Tab. 2
\begin{table}%[b]
\caption{\small Exponents $\alpha$ of $\langle \tau \rangle (N)$ and $\beta$ of $\sigma^2_{\tau}(N)$ from fits to Eqs.~(\ref{emoy}) and (\ref{var}), respectively, for all the elections in eight out of the nine countries shown in the upper part of Table~\ref{tdata}. $\alpha$ and $\beta$ are measured only in the range where Eqs.~(\ref{emoy}) and (\ref{var}) hold (see Figs.~\ref{ftau} and \ref{si_tau}). %Concerning Israeli municipalities with a Jewish majority, Eq.~(\ref{emoy}) holds with $\alpha\approx 0.32$ (see Fig.~\ref{israel}). 
For Polish elections  $\beta$ is not included in the average because of its very poor statistics.}
\begin{tabular}{l|llllllll|l}
\hline
\hline
& At & CR & Cz & Fr & Pl & Ro & Sk & Sp & average $\pm$ stdev \\ \hline
$\alpha \approx$ & 0.36 & 0.25 & 0.28 & 0.29 & 0.37 & 0.31 & 0.32 & 0.27 & 0.31 $\pm$ 0.04 \\ 
\hline
$\beta \approx$ & 0.34 & 0.21 & 0.41 & 0.33 &  & 0.26 & 0.35 & 0.34 & 0.32 $\pm$ 0.06 \\ 
\hline
\hline
\end{tabular}
\label{tstat}
\end{table}

It is worth noticing the difference between these results and those issued from previous studies of the LTR in the case of non-local elections for {\it the same constituencies}~\cite{diffusive2}  (shown in Fig.~\ref{si_tau}). The universal behavior of $\langle \tau \rangle (N)$ we just exhibited for local elections does not hold for national elections, where the curve may change not only from one country to another, but also from one election to another in the same country. Similar findings apply to the comparison between the variances, $\sigma^2_\tau (N)$,  of local and national elections though the curves are, as expected, noisier than for the first moment. Hence we conclude that the aforementioned regularities arise only when the purpose of the election {\it solely concerns the interests of the constituency (here, the municipalities)}.
Furthermore, our results go far beyond the confirmation of the well known fact that the electoral participation decreases with the constituency size (although there are exceptions to this rule for non-local elections, as shown in fig.~\ref{si_tau}). In particular we showed that, in the case of local elections, the decrease of $\langle \tau \rangle (N)$ and $\sigma^2_\tau (N)$ is the same, irrespective of the election date or the country considered.

\begin{figure}[t]
\centering
\includegraphics[width=7cm, height=5cm,clip=true]{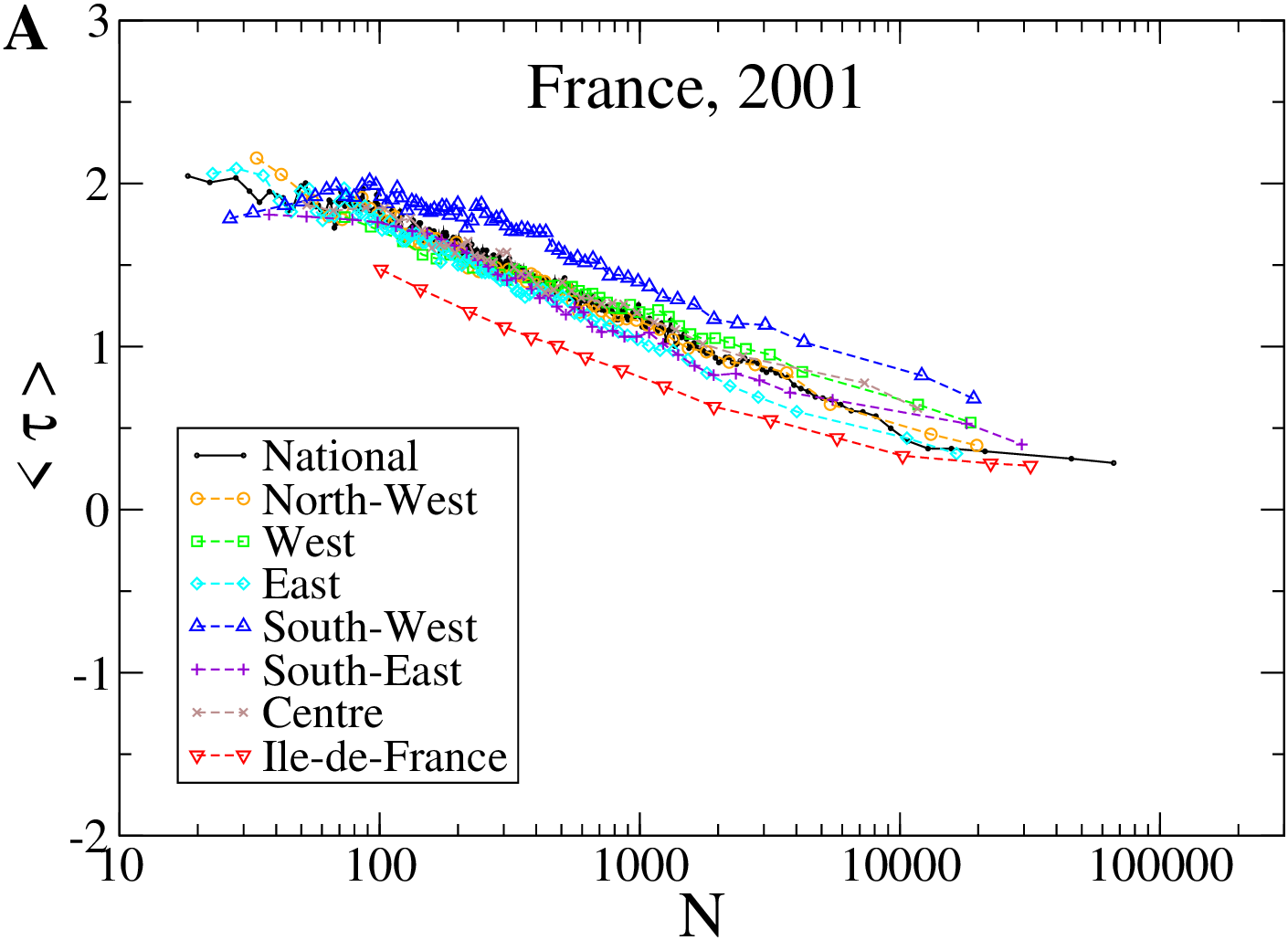}\\%{si-fr-reg-moy.eps} 
\includegraphics[width=7cm, height=5cm,clip=true]{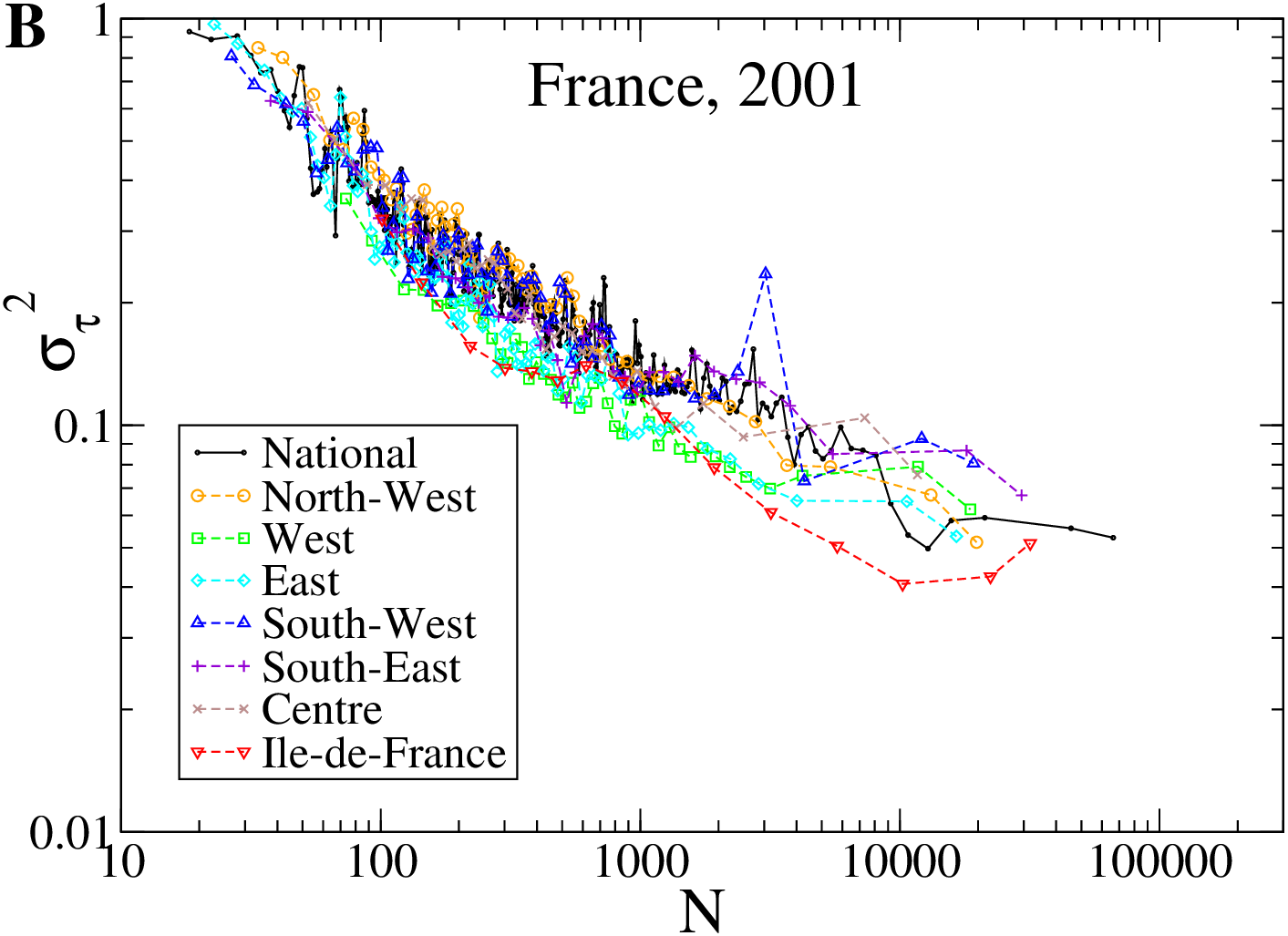}%{si-fr-reg-var.eps}
\caption{\small (Color online) ({\bf A})  Average, $\langle \tau \rangle (N)$, and ({\bf B}) variance, $\sigma^2_\tau (N)$, of the LTR corresponding to the 2001 local election,  plotted for each French region. Curves for all municipalities (Fr-2001), as in Fig.~\ref{ftau}, are also plotted for reference.}
\label{regions}
\end{figure}

\begin{figure}[t]
%\centering
\includegraphics[width=7cm, height=5cm,clip=true]{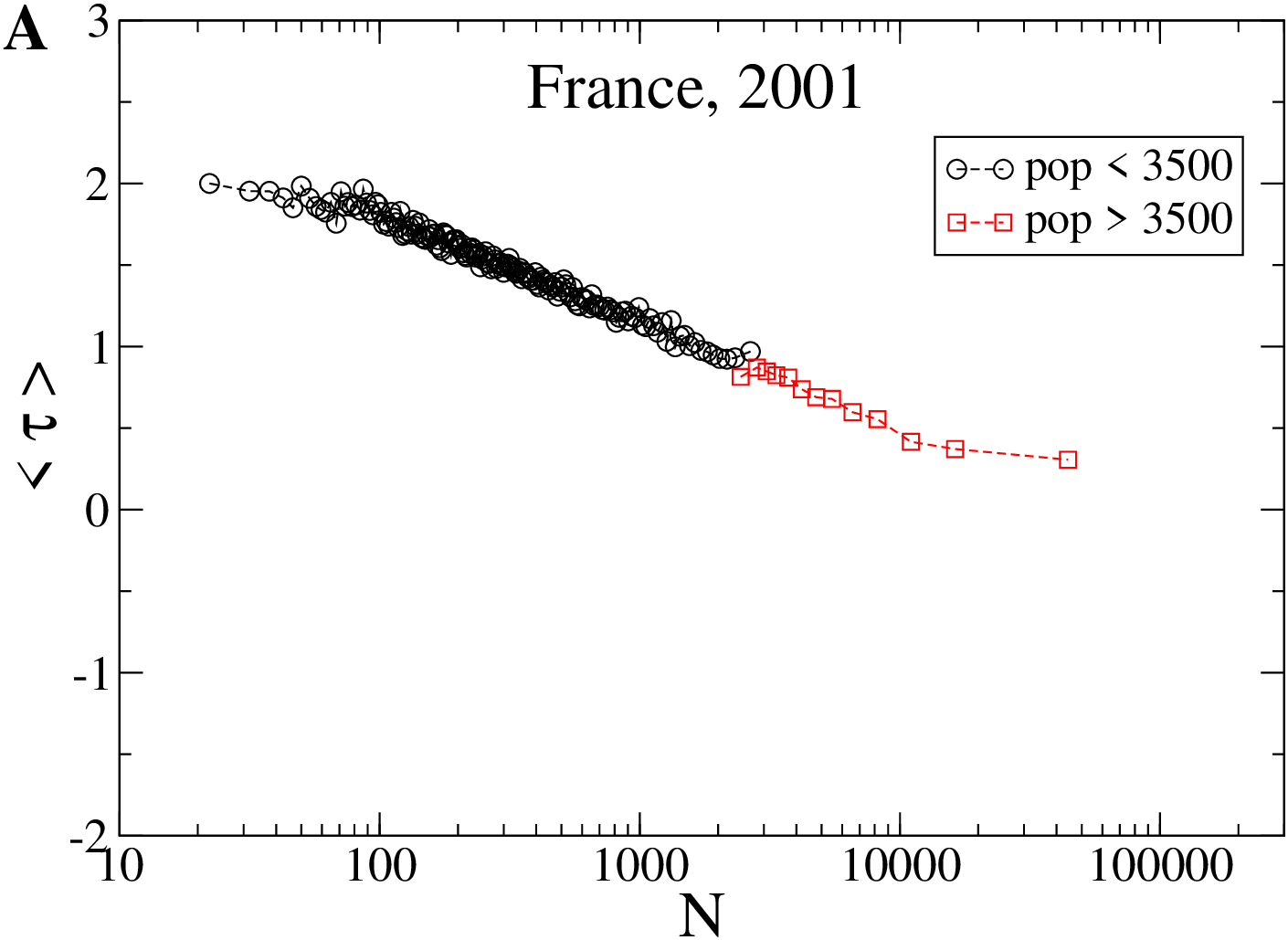}\\%{si-fr-2rules.eps}
\includegraphics[width=7cm, height=5cm,clip=true]{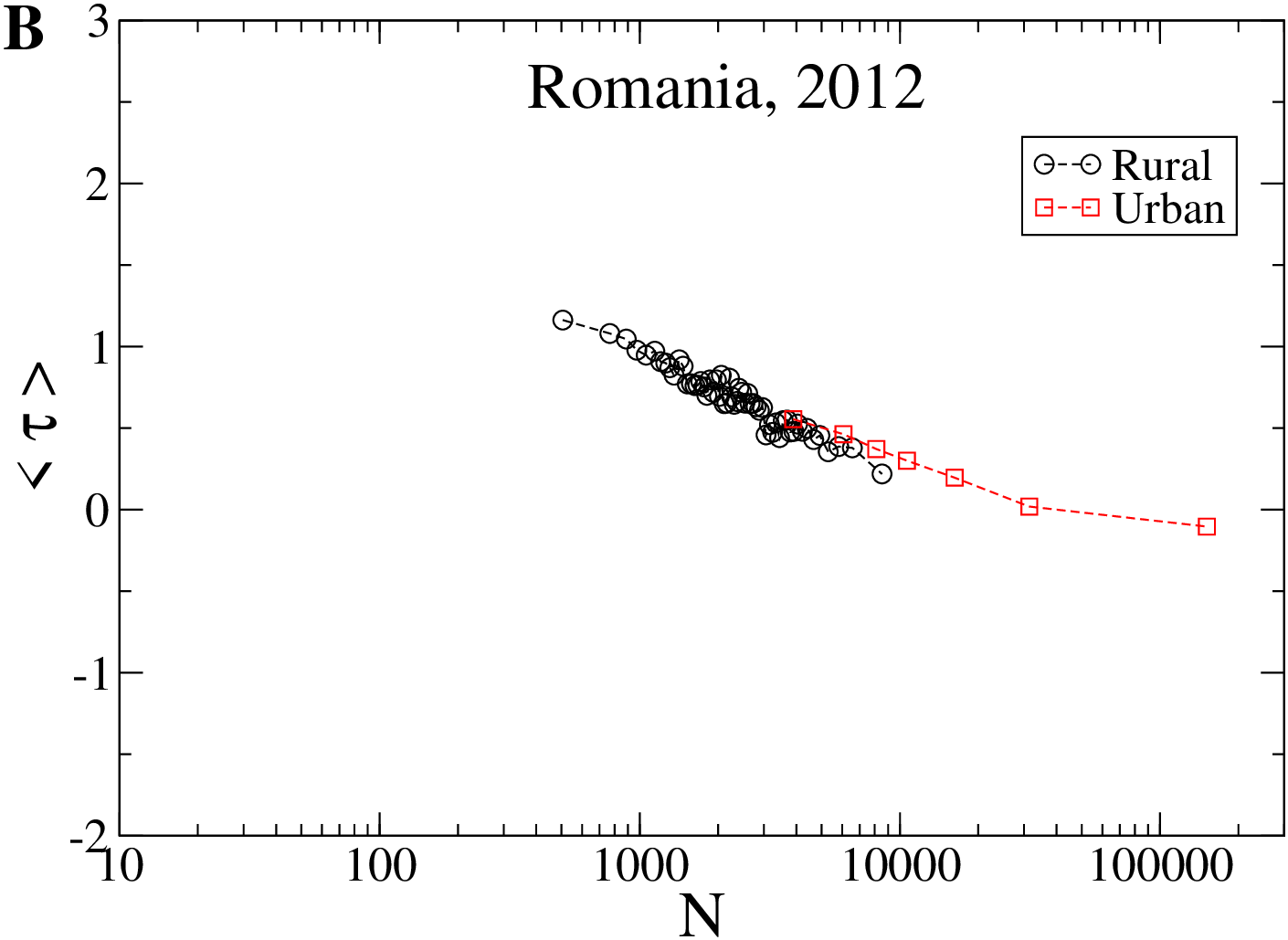}%{si-ro-rurb.eps}
\caption{\small (Color online) Robustness of the average LTR   $\langle \tau \rangle (N) $ behavior in local elections. ({\bf A}) Lack of influence of the voting rule, for French municipalities smaller and larger than the threshold limit of 3500 inhabitants. ({\bf B}) Lack of influence of the  rural or urban character of municipalities on the regularities observed  in local elections (2012 Romanian election).}
\label{robust_elec}
\end{figure}

We have studied the robustness of the observed behavior when data are grouped according to the different regions or provinces inside a given country. The results are shown in Fig.~\ref{regions} for French elections (we chose France because it has many more municipalities than other countries, leading to a better statistics). The data split into regions exhibit the same regularities as previously described for countries. For instance, the average $\langle \tau \rangle (N)$ for different regions follows the equation (\ref{emoy}), with the same slope $\alpha$ (and with different shifts $C$) that we obtained  when all the municipalities in the country were considered. 
Also, the results depend neither on the voting rules nor on the rural/urban character of the municipalities. In France, for instance, the voting rule is different according to the population of the municipality (the limit being 3500 inhabitants). Fig.~\ref{robust_elec}A shows that there is no kink in the slope of the line $\langle \tau \rangle (N)$ when the voting rule changes. Moreover, both curves fit the same line, thus indicating that the total turnout rate describes a property of that particular electoral event. We are able to separate rural from urban municipalities in Romanian data. As shown in Fig.~\ref{robust_elec}B, the behavior of the LTR is not different between both cases.
This result, we believe, strengthens our claim that the population size effect in local elections is universal.

%Fig. 3

%Fig. 4

To further check the robustness of our measures, we have evaluated the average of LTR (i.e., $\langle \tau \rangle (N)$) over municipalities with approximatively the same size $N$, in a different way. Instead of averaging $\tau$ over, for instance, 100 municipalities of size $\approx N$ as for $\langle \tau \rangle (N)$, we evaluate the LTR of an artificial town which results from aggregating data of these 100 municipalities.~\footnote{Let us consider 100 municipalities, $i$, of size $\approx N$; each one with a number of registered voters $N_i$ ($i=1,2,...,100$), a number of voters $N_{i,+}$ which leads to its LTR $\tau_i=\ln(\frac{N_{i,+}}{N_i-N_{i,+}})$. According to what precedes, $\langle \tau \rangle (N)=1/100\sum_{i=1}^{100}\tau_i$. We can define another kind of average LTR over these 100 municipalities by $\ln\left(\frac{\sum_{i=1}^{100} N_{i,+}}{\sum_{i=1}^{100} (N_i-N_{i,+})}\right)$.} The result obtained computing the LTR in this way are very similar to those described above (but the information on the variance is obviously lost).\\

\begin{figure}[t]
\includegraphics[width=6cm, height=4cm,clip=true]{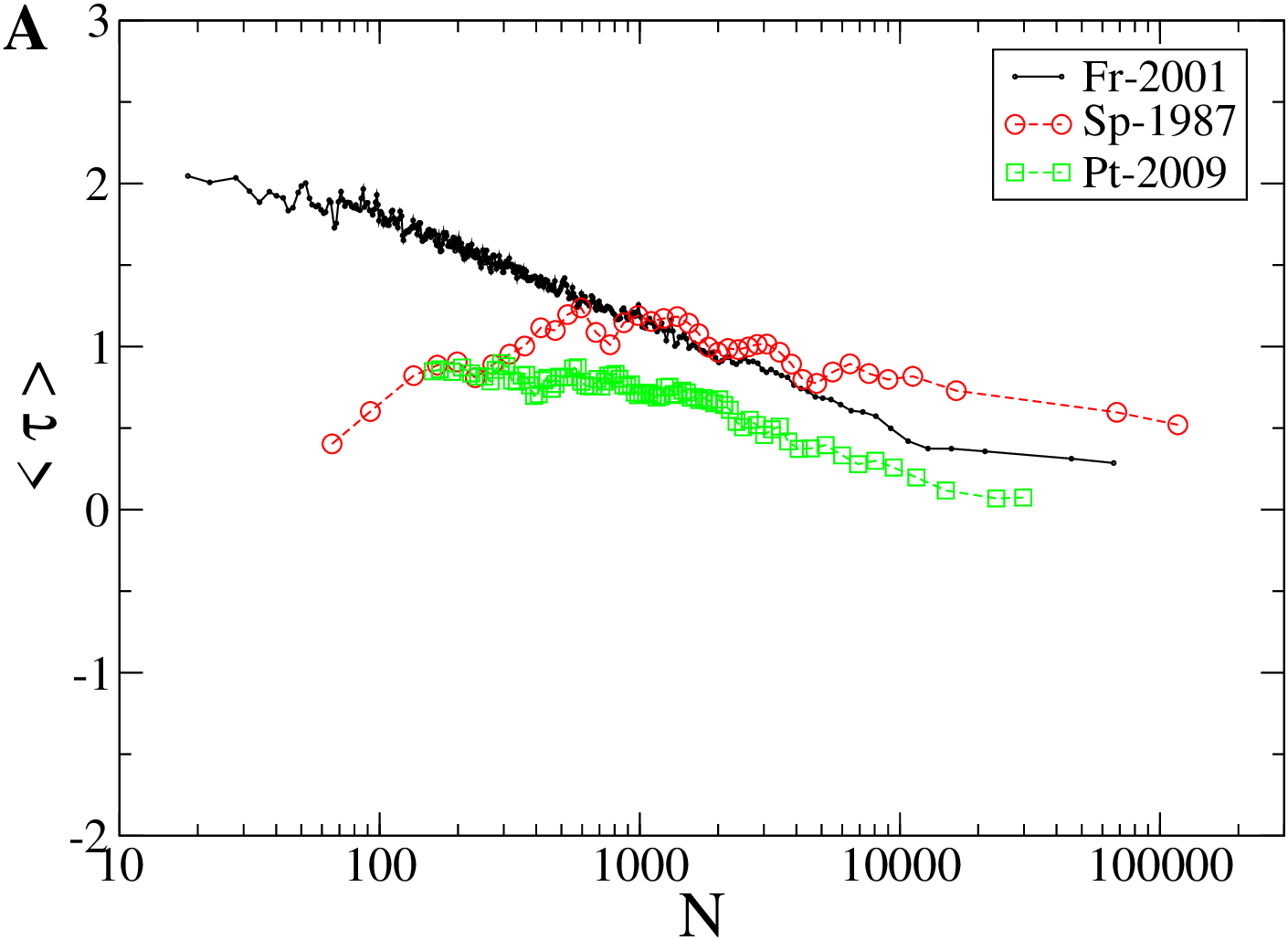}\\%{si-exceptions-moy.eps}
 \includegraphics[width=6cm, height=4cm,clip=true]{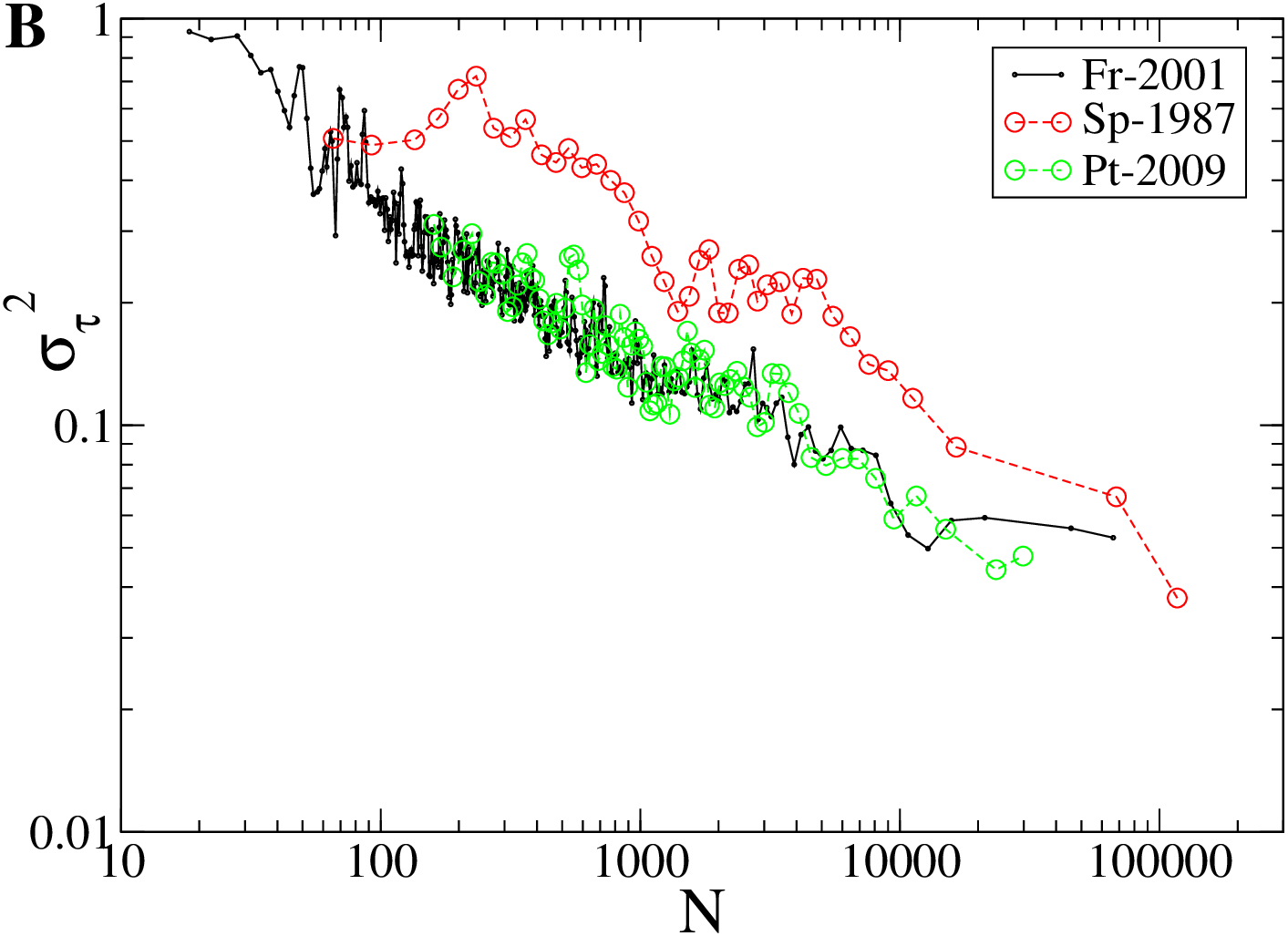}\\%{si-exceptions-var.eps} 
 \includegraphics[width=6cm, height=4cm,clip=true]{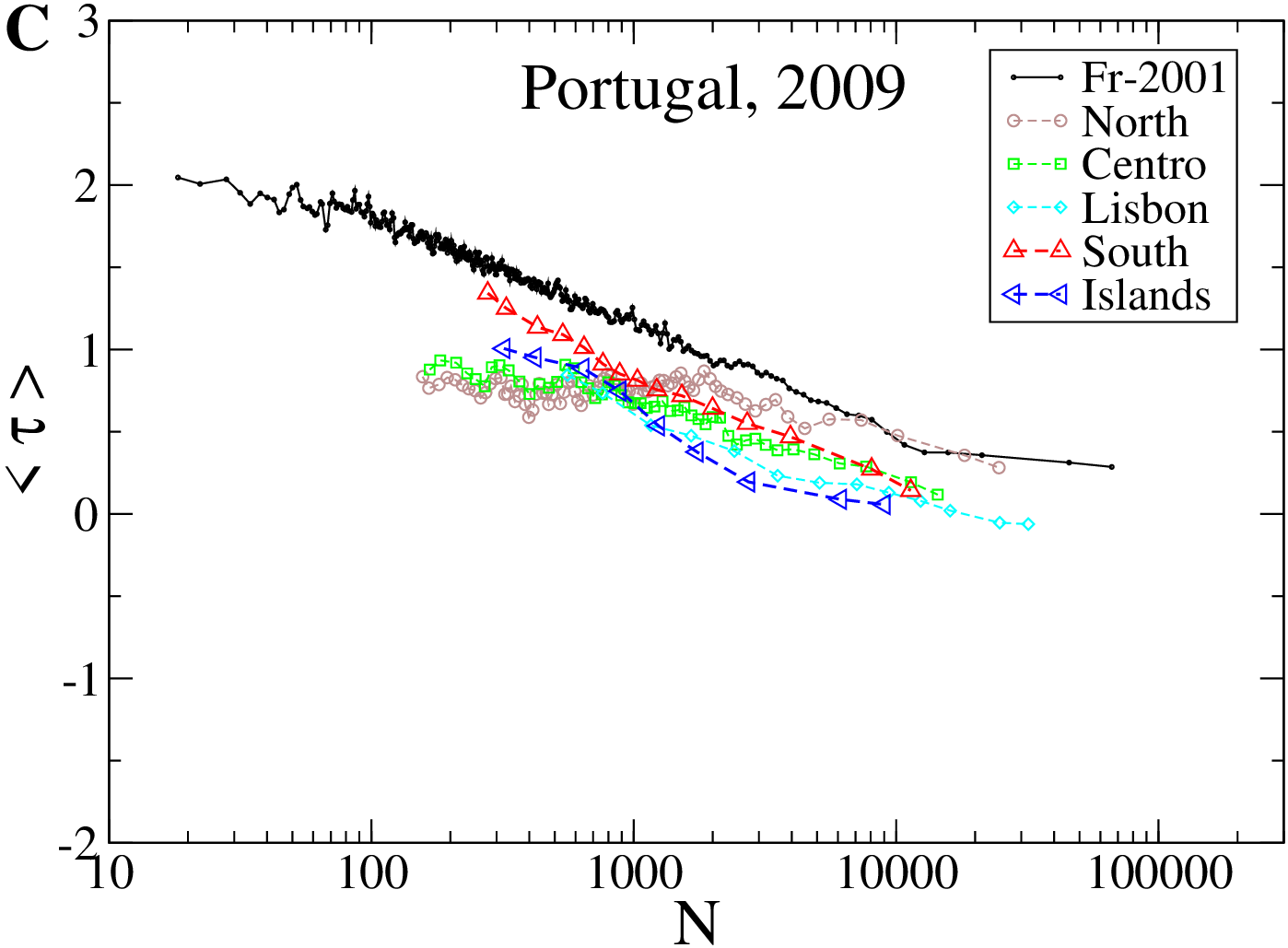}%{si-pt-reg.eps}
\caption{\small (Color online) Exceptions to the universal behavior of LTR; in all the panels we plot the 2001 local French election (Fr-2001) as a reference. ({\bf A} and {\bf B}) $\langle \tau \rangle (N)$ and $\sigma^2_\tau (N)$, respectively, for local elections in Portugal (Pt) and Spain (Sp). ({\bf C}) The Portuguese 2009 local election, different curves correspond to to municipalities of different regions.}
\label{exceptions}
\end{figure}

Let us now consider the cases where the described universal behavior does not hold (lower part of Table~\ref{tdata}). In Portuguese elections, a cutoff is imposed to large cities: they are divided into {\it freguesias} of $N\leq40000$. In this way, as large cities are divided, small or average constituencies are created, thus artificially modifying the number of the municipalities of a given size. This situation mainly concerns the province of {\it Lisboa}. However there are other provinces ({\it regi\~oes}) where there are very few municipalities that reach the threshold of $ N=40000$, and still these provinces (North and Centro) do not show the universal behavior of the average $\langle \tau \rangle (N)$ previously exhibited. At the same time, other provinces like {\it Alentejo} and {\it Algarve} or the two islands {\it A\c{c}ores} and {\it Madeira} show the aforementioned  regularities. Nevertheless, aggregating data at the level of the country (in order to have enough statistics), the $\sigma^2_\tau (N)$ behaves in agreement with Eq.~(\ref{var}), with the exponent $\beta\approx0.34$. We have neither an organizational nor a sociological or historical explanation for the single case of Spanish elections (1987) which is an exception to the observed universal behavior of the average LTR. However, we observe that $\sigma^2_\tau (N)$ behaves in agreement with Eq.~(\ref{var}), with the exponent $\beta \approx0.37$. These results are shown in Fig.~\ref{exceptions}.

%Fig. 5

\begin{figure}[t]
\includegraphics[width=7cm, height=5cm,clip=true]{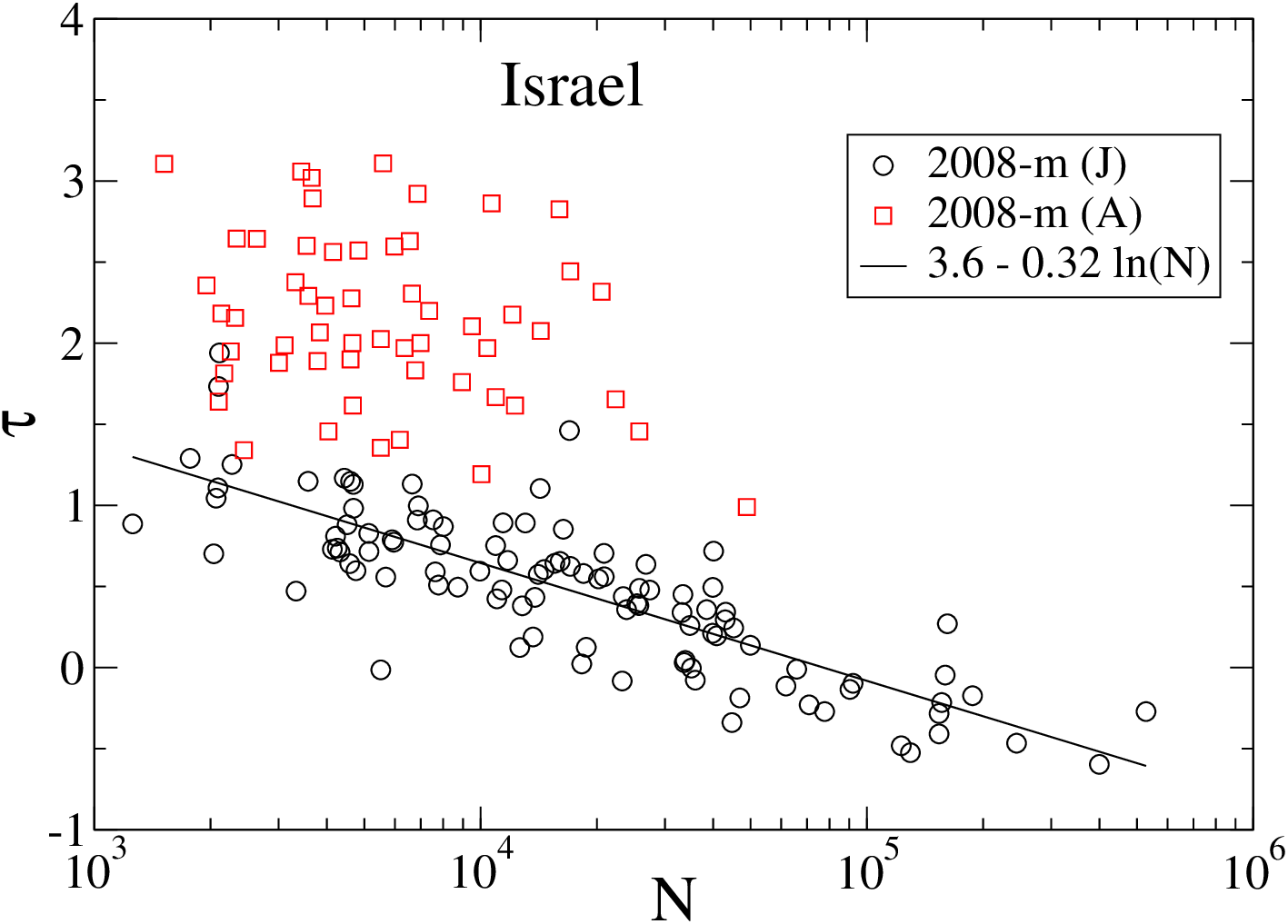}%{si-is.eps}
\caption{\small (Color online) Exceptions to the universal behavior of LTR. Scatter plot of the local elections in Israel; in red, municipalities with a majority of Arab population, in black, municipalities with a majority of Jewish population. The straight line corresponds to the fit of the black dots $\langle\tau\rangle (N) \approx 3.6-0.32\,\ln(N)$.}
\label{israel}
\end{figure}

The anomalous behavior of Israeli elections is particularly interesting. When plotting data of all the municipalities of the country, the behavior described by Eq.~(\ref{emoy}) is not observed, but as is shown in Fig.~\ref{israel}, when splitting data into municipalities with a majority of Jewish population on one side and with a majority of Arab population on the other, one retrieves the regularities in the first case but not in the second. This result reinforces the idea that the observed regularities reveal some property of the group.\\

%Fig. 6

Finally, we want to point out that the  LTR is, among other variables measuring vote participation, particularly interesting when it comes to exhibiting regularities. Indeed, the average value of the {\it turnout rate} $p$ also shows a linear decay with $log(N)$ as in Eq.~(\ref{emoy}), but the slope is not universal. Moreover, the variance of $p$ (the equivalent for $p$ of $\sigma^2_\tau (N)$) does not exhibit any regularity when plotted as a function of $N$. One possible reason for this, is that unlike $\tau$, $p$ is bounded ($p\in [0,1]$). Other quantities like the {\it involvement entropy}~\cite{weak-law} do not show clear quantitative regularities.\\

We would now like to study \emph{how} the constituency size influences the LTR. We begin by noting that Eq.~(\ref{var}) indicates that the LTR cannot be understood as resulting from a process where the voters act independently (in such a case one must find $\sigma^2 \approx 1 / N$). On the contrary, it is compatible with a subdivision of the constituency of size $N$ into $n_g \approx N^{\beta}$ independent groups. In order to further investigate this hypothesis, we study how the number of democratic representatives of a given constituency scales with its size. The intuition is that, if significant, this group structure must be mirrored in the political organization at different scales. 
It is also interesting to note at this point that from the fit of Eqs.~(\ref{emoy}) and (\ref{var}),  $\alpha \approx \beta$. In this case, Eq.~(\ref{emoy}) reads  $\mathrm{d}\langle\tau\rangle \approx -\frac{\mathrm{d}n_g}{n_g}$, consistent with  $n_g=N^\beta$. This point will be used later in the formulation of a phenomenological model.

%%%%%%%%%%%%%%%%%%%%%%%%%%%%%%%%%
\section{Number of democratic representatives}

We have gathered and analyzed data regarding the number of representatives, $n_r$, of different democratic assemblies  as a function of the  constituency size, for  different countries and  scales (municipality, state/region/province, country). We consider only elections where the representatives are directly elected by the citizens, and for states or countries with bicameral legislatures, only the lower house (or House of Representatives) is considered (see details in the Appendix). Fig.~\ref{frepres} shows the dependence of the number of the representatives on the size of the constituency at national and municipal scales.

%Fig7
\begin{figure}[t]
\includegraphics[width=8cm, height=6cm, clip=true]{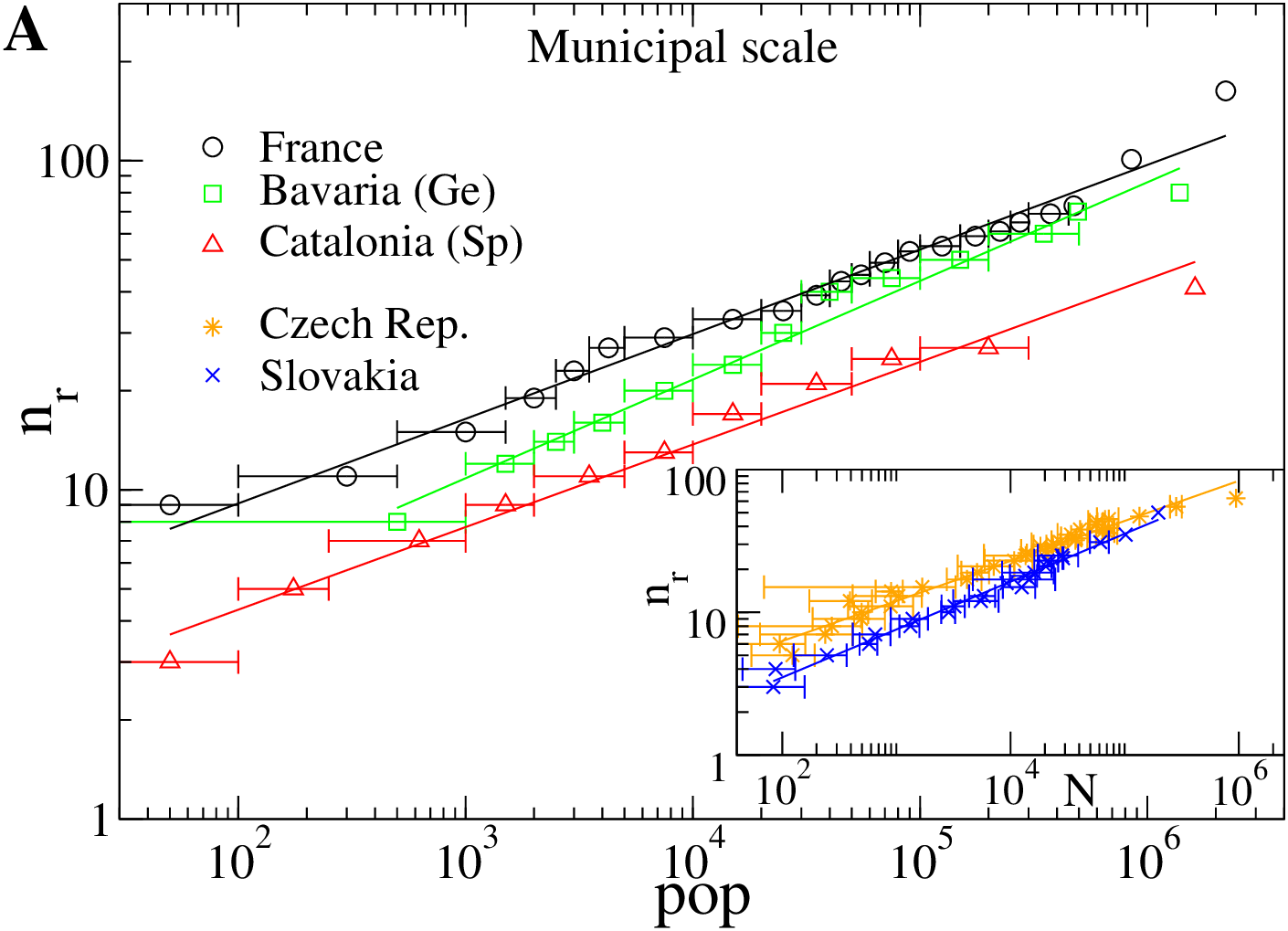}\\%[width=7cm, height=5cm, clip=true]
\includegraphics[width=8cm, height=6cm, clip=true]{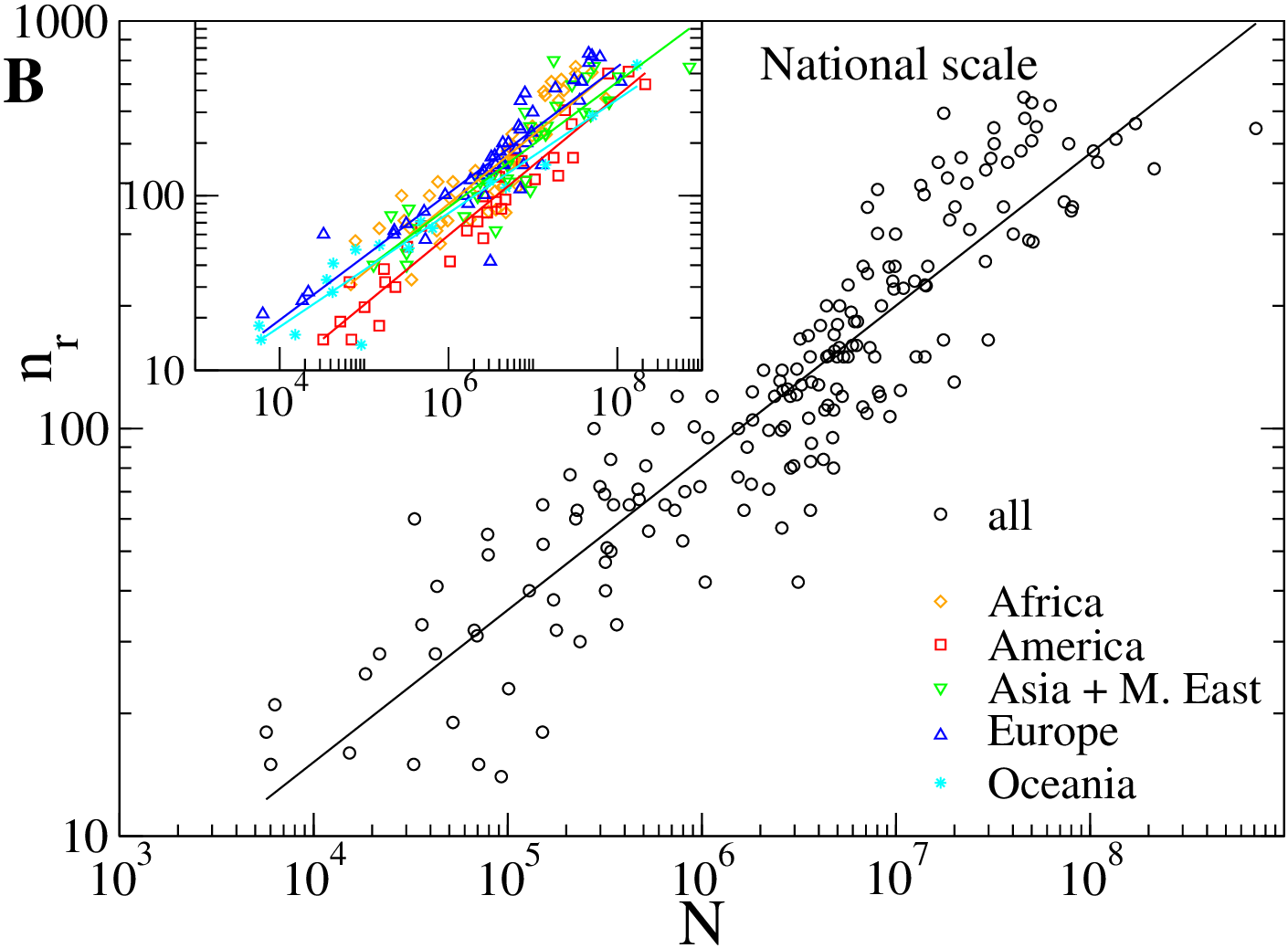}
\caption{\small (Color online) Log-log plot of the total number of democratic representatives, $n_r$, at municipal and national scales as a function of the total number of registered voters, $N$, or the population size, $pop$. ({\bf A}) Municipal councils. For the three cases of the main panel, plotted as a function of the population size, the electoral regulations impose that all the cities with a population between two given values have the same number of representatives. The error bars represent the corresponding interval. In the inset we plot the number of representatives in municipal assemblies in Slovakia and the Czech Republic as a function of the registered voters. In these cases the corresponding number is given city by city. The points represent the average value of the number of registered voters $N$ of the  cities having the same $n_r$ and the bars the standard deviation. ({\bf B}) National chambers. In the main panel, number of representatives as a function of $N$ of all the considered countries. In the inset, the same countries appear grouped by continents.}
\label{frepres}
\end{figure}

A power law  correctly fits the data over five decades, giving:  
\begin{equation}
n_r \sim E\,N^{\gamma}\,,
\label{eq_repr}
\end{equation}
with  $E$ being a chamber-dependent constant of the order of the unity. Again, the exponent is  $\gamma \approx 1/3$. Tab.~\ref{trepres} gives the details of the fitting parameters. 

\begin{figure}[t]
\includegraphics[width=4.4cm, height=3.7cm, clip=true]{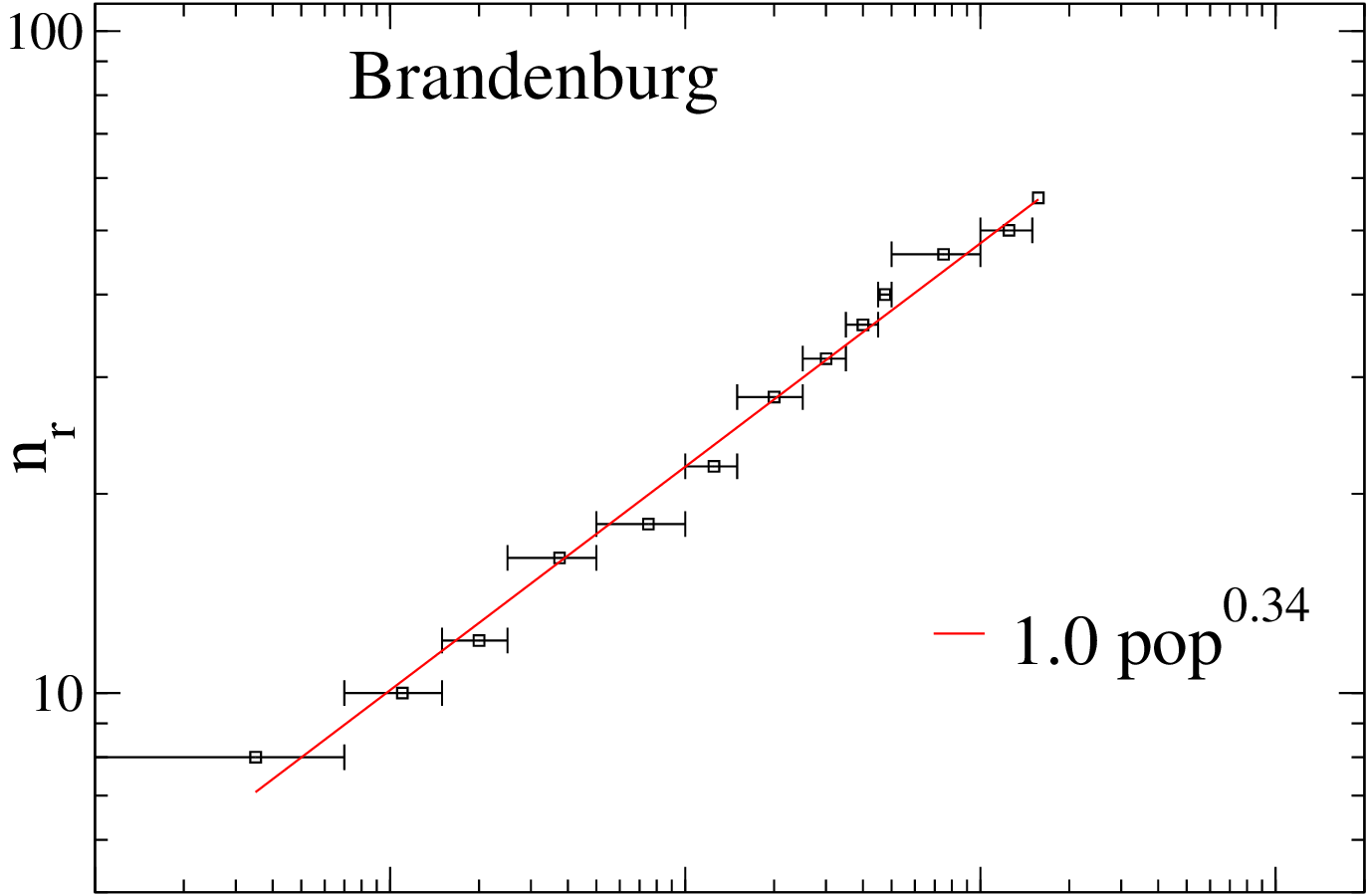}
\includegraphics[width=4cm, height=3.7cm, clip=true]{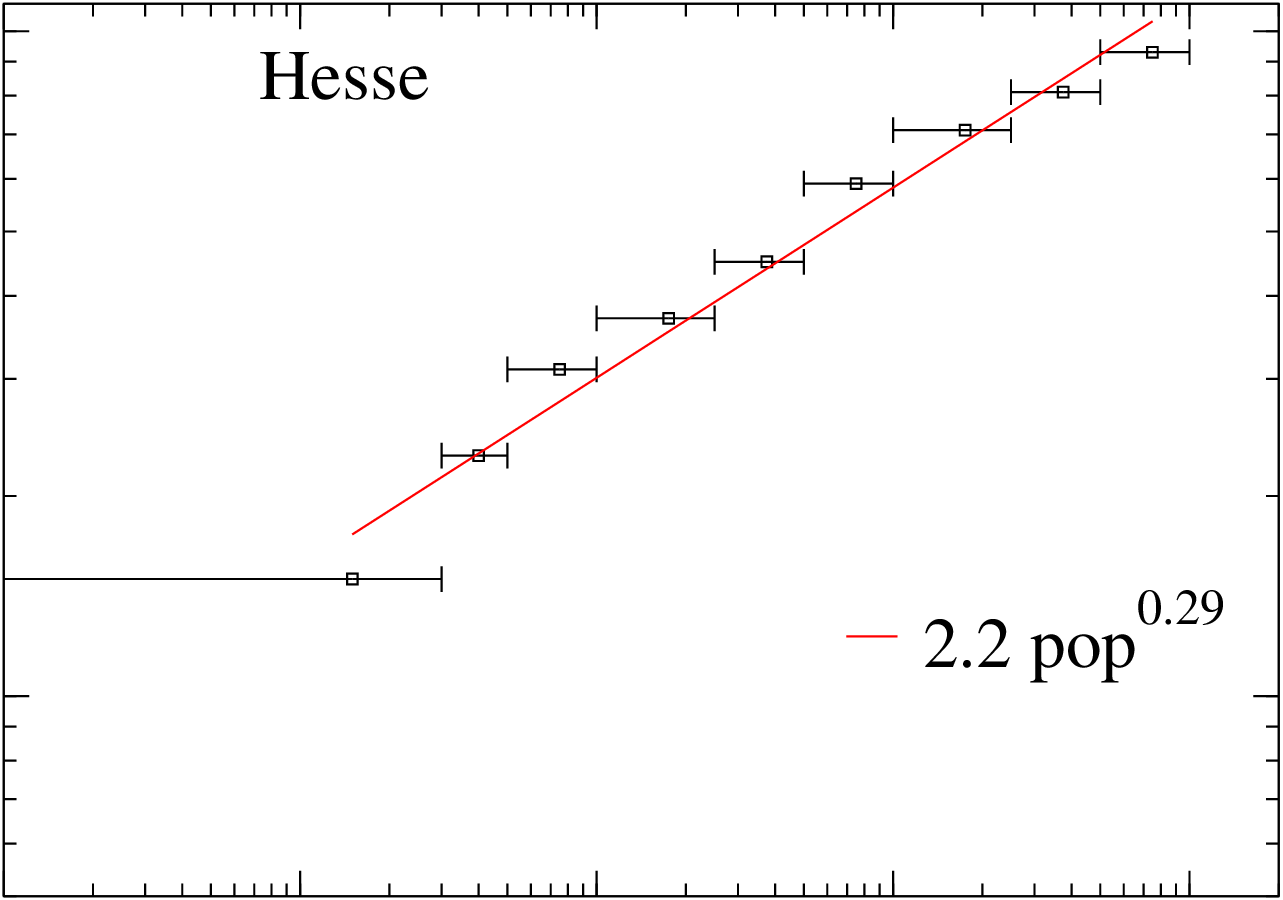}\\
\includegraphics[width=4.4cm, height=3.7cm, clip=true]{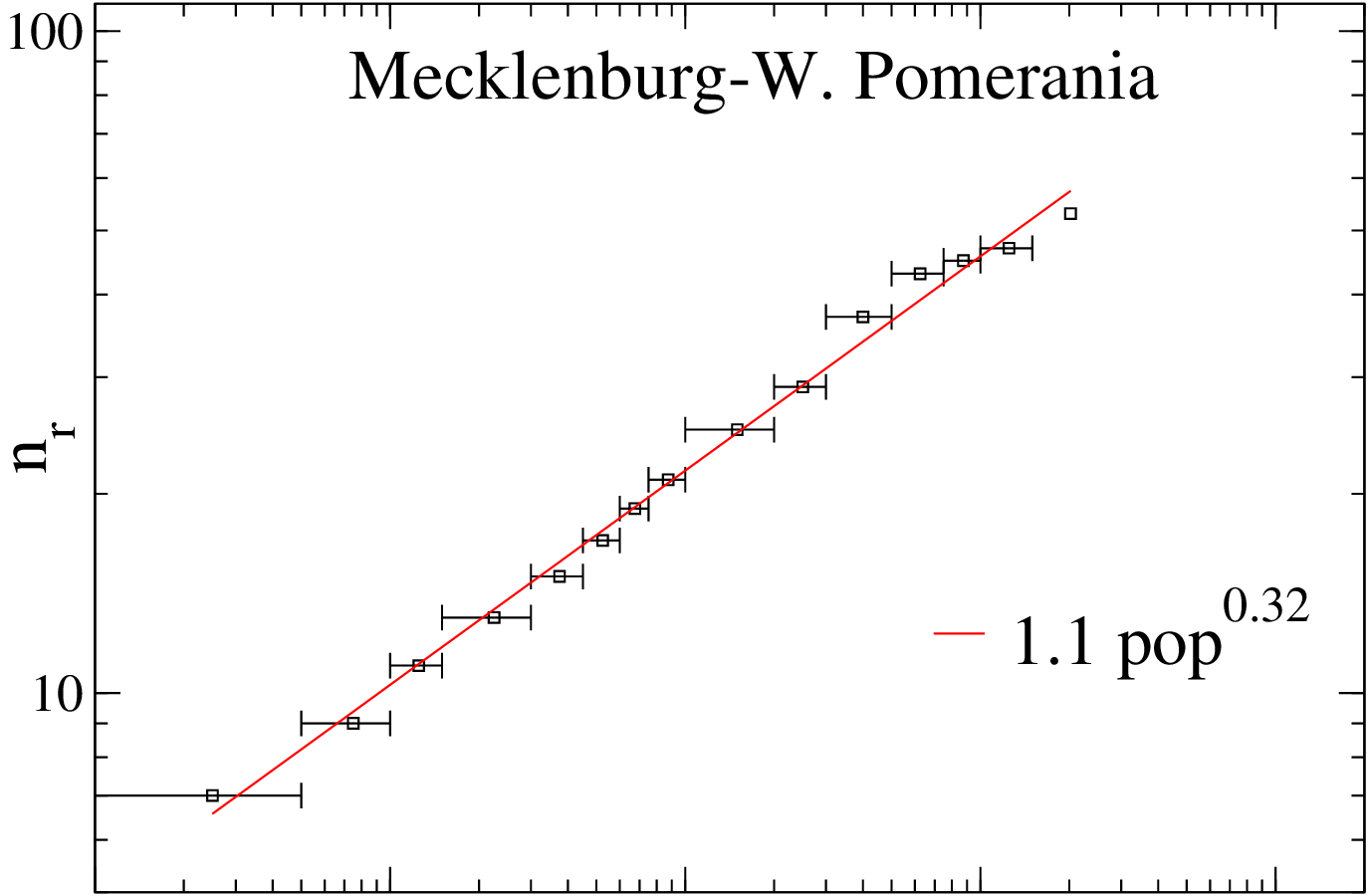}
\includegraphics[width=4cm, height=3.7cm, clip=true]{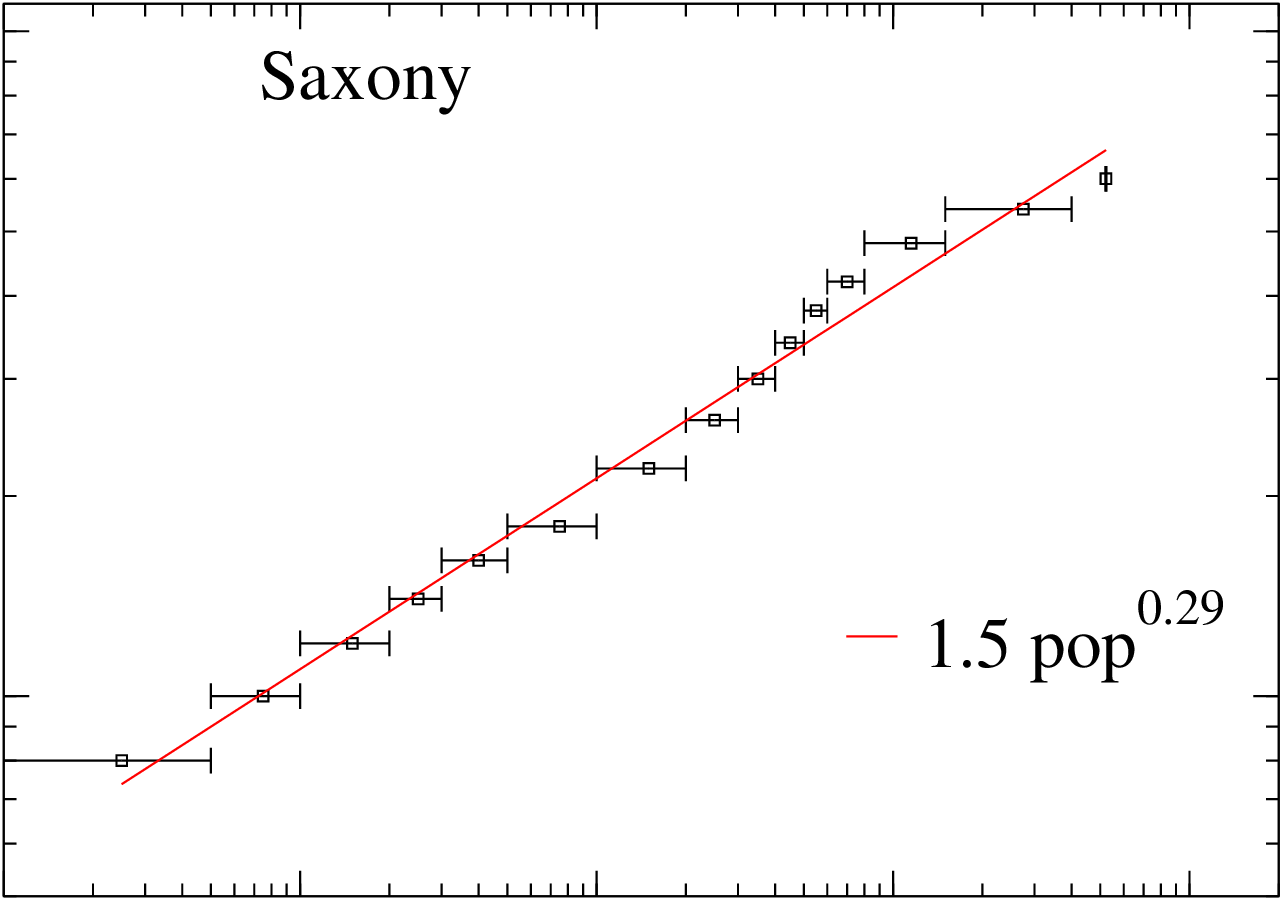}\\
\includegraphics[width=4.4cm, height=4.2cm, clip=true]{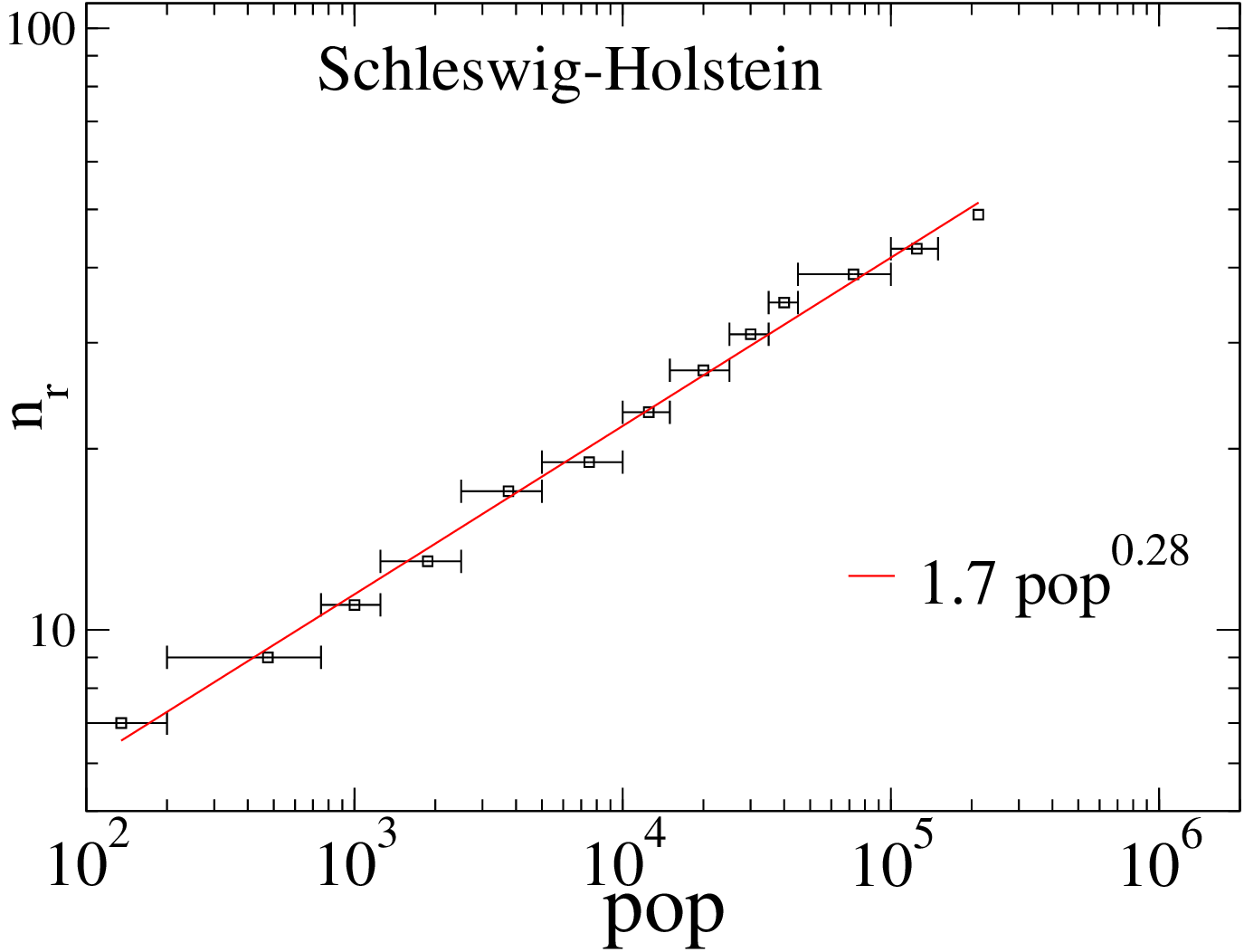}
\includegraphics[width=4cm, height=4.2cm, clip=true]{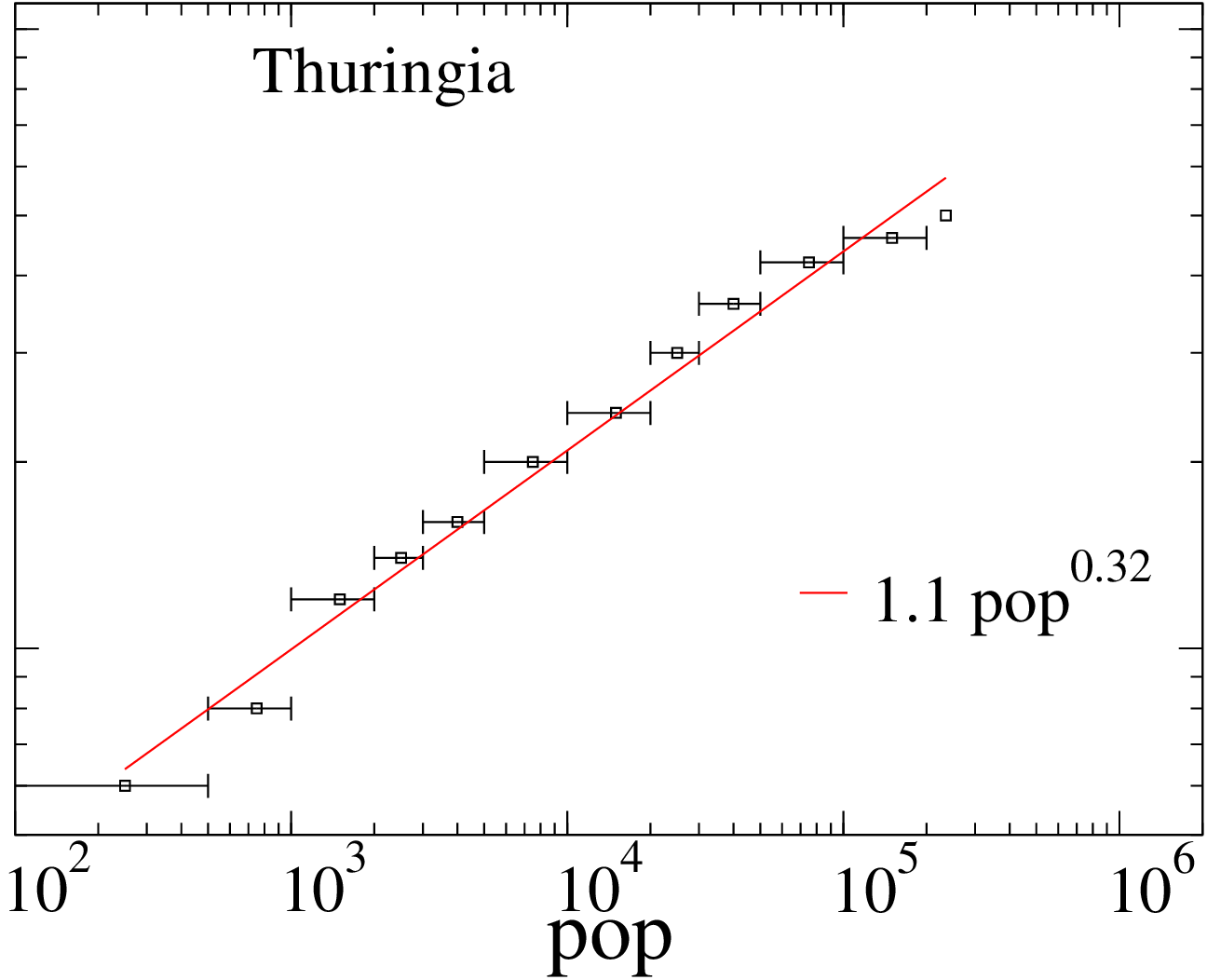}
\caption{\small (Color online) Number of representatives to town councils in different German {\it L\"ander}. The electoral rule fixing the number of representatives as a function of the population is different for each {\it Land}. The case of Bavaria (the one spreading the largest range of constituency sizes) is plotted in Fig.~\ref{frepres}A.}
\label{fsi-repres-ge}
\end{figure}

The case of Germany is particularly interesting because the behavior described by Eq.~(\ref{eq_repr}) still holds (see Fig.~\ref{fsi-repres-ge}), despite the fact that the rules fixing the number of representatives to municipal councils are different from one {\it Land} to another.

%Fig. 8

%Fig. 9
\begin{figure}[t]
\includegraphics[width=7cm, height=5cm,clip=true]{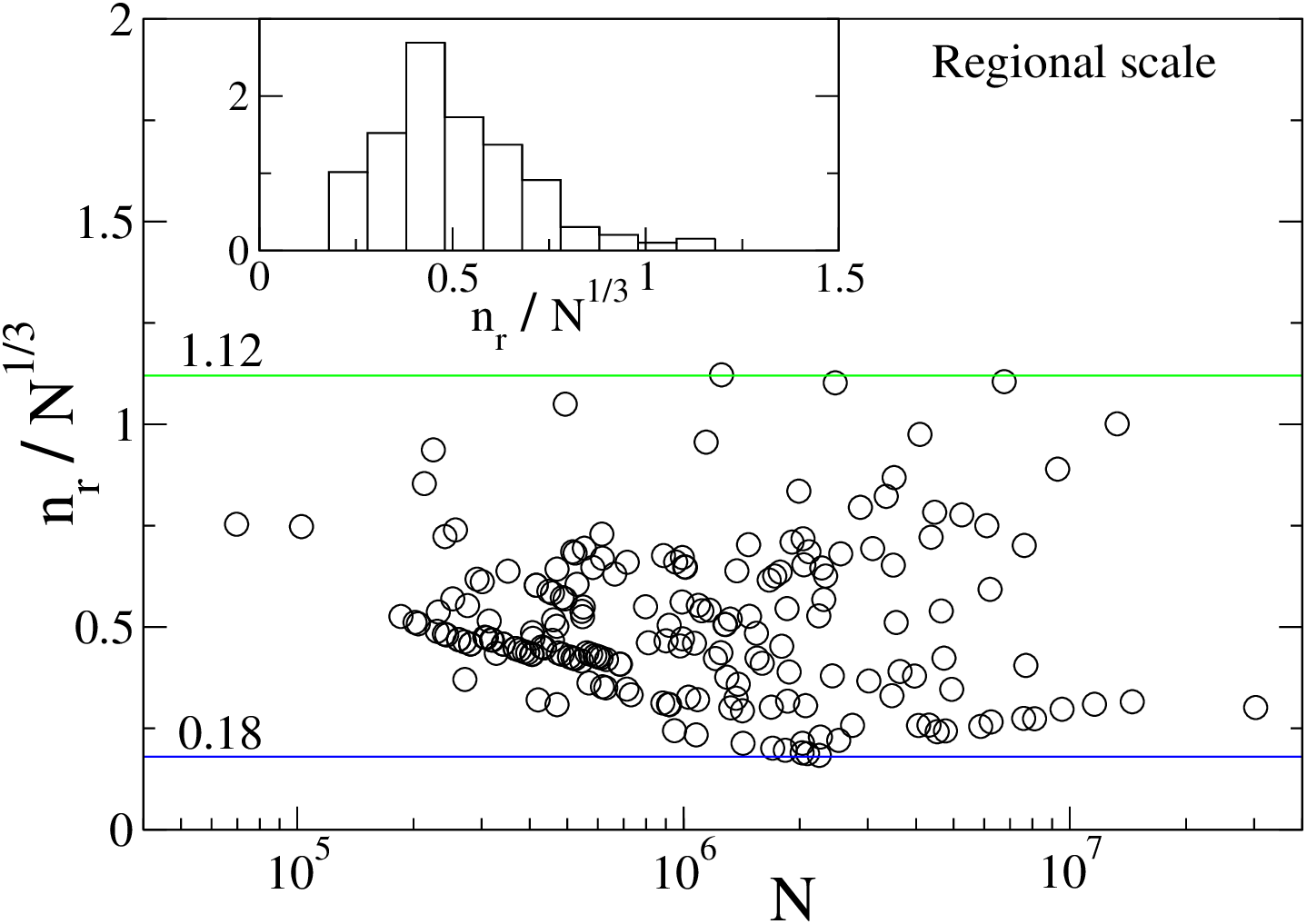}%{ratio-reg.eps} 
\caption{\small (Color online) Prefactor $E=n_r/N^{1/3}$ of the power-law fit of $n_r$ with $\gamma=1/3$ (see Eq.~(\ref{eq_repr})), for 197 Regional/States Chambers in 11 countries: nine in Austria, 27 in Brazil, 13 in Czech Republic, 26 in France, 16 in Germany, 21 in Italy, 16 in Poland, two in Portugal, 42 in Romania, eight in Slovakia, and 17 in Spain. Inset: Histogram of the prefactor $E$ for these Chambers.} 
\label{fratio-reg}
\end{figure}

These results hold for representative assemblies at all scales. For the intermediate scale (called regions, states, or provinces, according to the country) data are extended over fewer than two decades which is too small an interval to fit a power law with good confidence. Nevertheless, if we do so the results of the fit are compatible with those found for  other scales. We alternatively show the ratio $n_r/N^{1/3}$, which corresponds to the prefactor $E$ of Eq.~\ref{eq_repr} in the case where $\gamma=1/3$, on Fig.~\ref{fratio-reg}. This quantity lies in the range $0.2 \lesssim E \lesssim 1.1$.

\begin{figure}[t]
\includegraphics[width=7cm, height=5cm,clip=true]{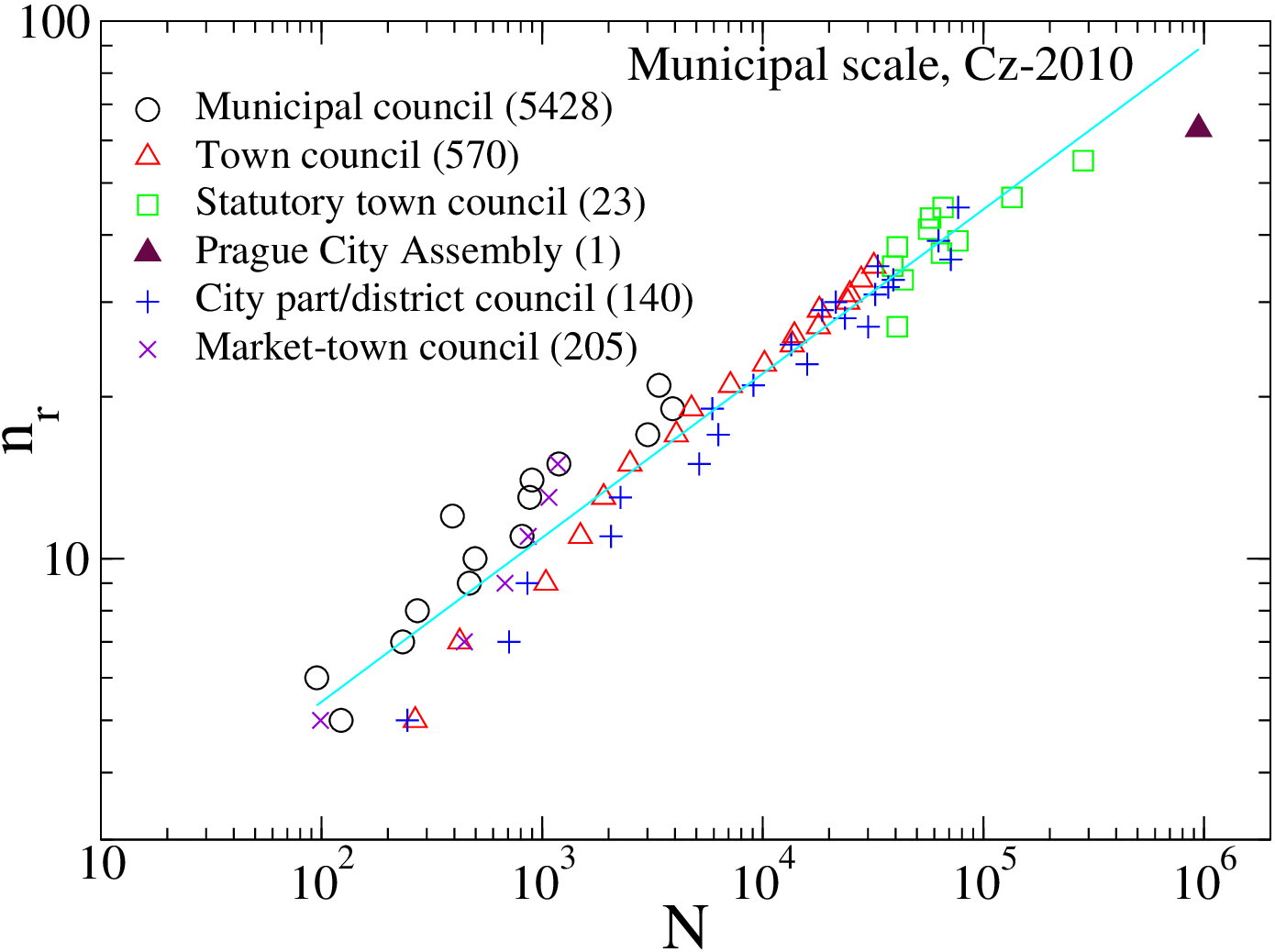}%{repres-mun-cz.eps} 
\caption{\small (Color online) Number of councillors at municipal scale for different kind of assemblies in the Czech Republic. In brackets, the number of councils of each kind. In this case the corresponding number of representatives is given city by city; the points represent the average number of registered voters of the considered cities that have the same number of representatives. The line corresponds to the power-law fit  of all the points.}
\label{multiple_assembly}
\end{figure}

It is interesting to note that the behavior given by Eq.~(\ref{eq_repr}), does not depend on the particular type of democratic chamber even at a given scale. For instance, in the Czech Republic, different democratic assemblies exists at the municipal scale: Municipal Council, Town Council, Statutory Town Council, Prague City Assembly, City Part/District Council, Market Town Council, etc. The results for these assemblies are shown in Fig.~\ref{multiple_assembly}.

%Fig. 10

In order to investigate the behavior of $n_r$ as a function of other extensive variables characterizing the studied population, we have plotted the $n_r$ of national assemblies as a function of surface of the country $S$, and the nominal gross domestic product (GDP). The results shown in Table~\ref{si-trepres} confirm that, among the variables studied here, the best fit to a power law is obtained as a function of $N$.

%Tab. 3
\begin{table*}
\caption{\small Power-law fits of $n_r$ with respect to $N$ at municipal and national scales (for municipal representatives in France, Catalonia and German {\it L\"ander}, data are given in terms of the population, $pop$, instead). The goodness of fit for municipal representatives is in each case $R^{2} > 0.97$ ($R^2$ values for national ones are indicated in  Tab.~\ref{si-trepres}).}
\begin{tabular}{ l | l l | l l | l l }
\hline
\hline
Municipal scale & France & $2.8\;pop^{0.26}$ & Catalonia (Sp) & $1.4\;pop^{0.25}$ & Bavaria (Ge) & $1.4\;pop^{0.30}$\\
 & Brandenburg (Ge) & $1.0\; pop^{0.34}$ & Hesse (Ge) & $2.2\; pop^{0.29}$ & Mecklenburg-W. Pomer. (Ge) & $1.1\; pop^{0.32}$\\
 & Saxony (Ge) & $1.5\; pop^{0.29}$ & Schleswig-Holstein (Ge) & $1.7\; pop^{0.28}$ & Thuringia (Ge) & $1.1\; pop^{0.32}$\\
 & Slovakia & $0.75\,N^{0.34}$ & Czech Republic & $1.7\,N^{0.28}$ & & \\
\hline
National scale & all & $0.49\,N^{0.37}$ & Africa & $0.39\,N^{0.39}$ & America & $0.24\,N^{0.40}$\\
 & Asia \& M.East & $0.60\,N^{0.36}$ & Europe & $0.69\,N^{0.36}$ & Oceania & $0.89\,N^{0.33}$\\
\hline
\hline
\end{tabular}
\label{trepres}
\end{table*}

%Tab. 4
\begin{table*}
\caption{\small Power-law fits of the total number of deputies with respect to an extensive quantity of the considered country. Parameters of the fit are written on the left, while the goodness of fit $R^2$ is on the right. The extensive quantities are the number of registered voters ($N$), the total area ($S$ in $\mathrm{km^2}$), and the nominal gross domestic product ($GDP$, in billion US~\$). The total number of considered countries per continent or for the world (all) is given in parentheses.}
\begin{tabular}{ l | l r | l r | l r | l r | l r | l r  }
%\cline{2-13}
\hline
\hline
\multicolumn{1}{c|}{} & \multicolumn{2}{|c|}{all (181)} & \multicolumn{2}{|c|}{Africa (50)} & \multicolumn{2}{|c|}{America (34)} & \multicolumn{2}{|c|}{Asia \& M.E. (34)} & \multicolumn{2}{|c|}{Europe (46)} & \multicolumn{2}{|c}{Oceania (17)}\\
 \hline
$N$ & $0.49\, N^{0.37}$ & 0.84 & $0.39\, N^{0.39}$ & 0.79 & $0.24\, N^{0.40}$ & 0.93 & $0.60\, N^{0.36}$ & 0.81 & $0.69\, N^{0.36}$ & 0.84 & $0.89\, N^{0.33}$ & 0.91\\
$S$ & $7.3\, S^{0.25}$ & 0.60 & $6.6\, S^{0.26}$ & 0.56 & $3.7\, S^{0.28}$ & 0.84 & $18\, S^{0.19}$ & 0.28 & $11\, S^{0.25}$ & 0.63 & $6.1\, S^{0.25}$ & 0.80\\
$GDP$ & $51\, GDP^{0.26}$ & 0.48 & $97\, GDP^{0.14}$ & 0.11 & $28\, GDP^{0.31}$ & 0.80 & $67\, GDP^{0.22}$ & 0.26 & $45\, GDP^{0.29}$ & 0.47 & $44\, GDP^{0.28}$ & 0.83\\
\hline
\hline
\end{tabular}
\label{si-trepres}
\end{table*}

This power-law dependence can be easily obtained using the following heuristic, which is reminiscent of the {\it least difficulty principle}~\cite{toulouse}. A good representative democracy must meet two competing requirements. First, it must maximize the representativeness, that is to say, the number $n_r$ of representatives should be such that the average number of citizens per representative $N_1 = N/n_r$ is as small as possible. On the other hand, it must be efficient: in order to make decisions a reasonable number of interactions among the $n_r$ representatives should take place. The higher this number of interactions, the more difficult it is to make a decision. Roughly, if one considers all pairwise interactions, the corresponding term is of the order of $n_r^2$. Therefore, a `good' representative regime is such that the function: $\mathcal{F} = N_1 + A\,n_r^2 $, where $A$ is a measure of the relative importance of representativeness and efficiency, is minimum. This leads immediately to $n_r \approx N^{1/3}$, which is in the range of the observed values.

A short historical investigation of the behavior of the number of representatives with the size of the respective constituencies in old democratic assemblies can be found in the Appendix, which in most of the cases do not correspond to the behavior described here. One reason for this discrepancy could be the fact that those ancient assemblies did not correspond to a democratic regime as we understand it nowadays. 

The evolution of the number of  representatives of {\em one assembly} with time is possible in the case of the United States, where an important increase of the  population (compared to European democracies) occurs along with a stable democratic regime. The results obtained (see the Appendix) are compatible with the behavior of the $n_r$  described here.

%%%%%%%%%%%%%%%%%%%%%%%%%%%%%%%%%%%%%%
\section{Phenomenological model for local elections}

The main idea behind the following phenomenological model is quite simple:  when facing a decision, $N$ agents behave as if they were grouped into $n_g \approx N^{\delta}$ subgroups, with $\delta \approx 1/3$. This idea is suggested by the observed organization of constituencies into groups, which holds for different scales, and the fact that $\alpha \approx \beta \approx \gamma \approx 1/3$. We do not make any assumption about the exact nature or the origin of these groups; this problem lies beyond the scope of this study. We just assume that people belonging to a given subgroup share some common traits. This organization in groups is assumed to go on at all scales, reflecting a hierarchy of differences among agents (as the subdivisions go on, agents inside the same group are more and more similar one to another).

Our assumptions relate to the problem of vote participation in local elections in the following way. Let $x$ be a random variable characterizing the propensity of one agent to go to vote to a particular election. $x$  is the sum of two independent contributions: the first one, $\zeta$ (i.i.d. with variance $\sigma^2_\zeta$), characterizes the propensity of people belonging to the same group, and the one, $\epsilon$, represents the variability of the individual. $\epsilon$ may fluctuate around the average value of the group's propensity, with variance $\sigma^2_\epsilon$. Hence, an agent $i$ who belongs to the group $g_i$ is characterized by a propensity to vote: $x_i = \zeta_{g_i} + \epsilon_i$. Assuming that there are $N$ agents distributed in $n_g$ groups of approximatively the same size, the variance of $\Sigma=\frac{1}{N}\sum_{i=1}^N x_i$ (representing the total propensity to vote) is $Var(\Sigma) = \frac{1}{n_g}\sigma^2_\zeta + \frac{1}{N}\sigma^2_\epsilon$. If $n_g \approx  N^{\delta}$ the previous equation yields, for $N\gg 1$ and for $\sigma^2_\zeta$ and $\sigma^2_\epsilon$ of the same order magnitude: $Var(\Sigma)\approx \frac{1}{n_g}\sigma^2_\zeta$, which is the dependency that we observe in data with $\delta \approx 1/3$. In what follows we will take $\delta = 1/3$. 

\begin{figure}[t]
\includegraphics[width=8cm, height=6.5cm, clip=true]{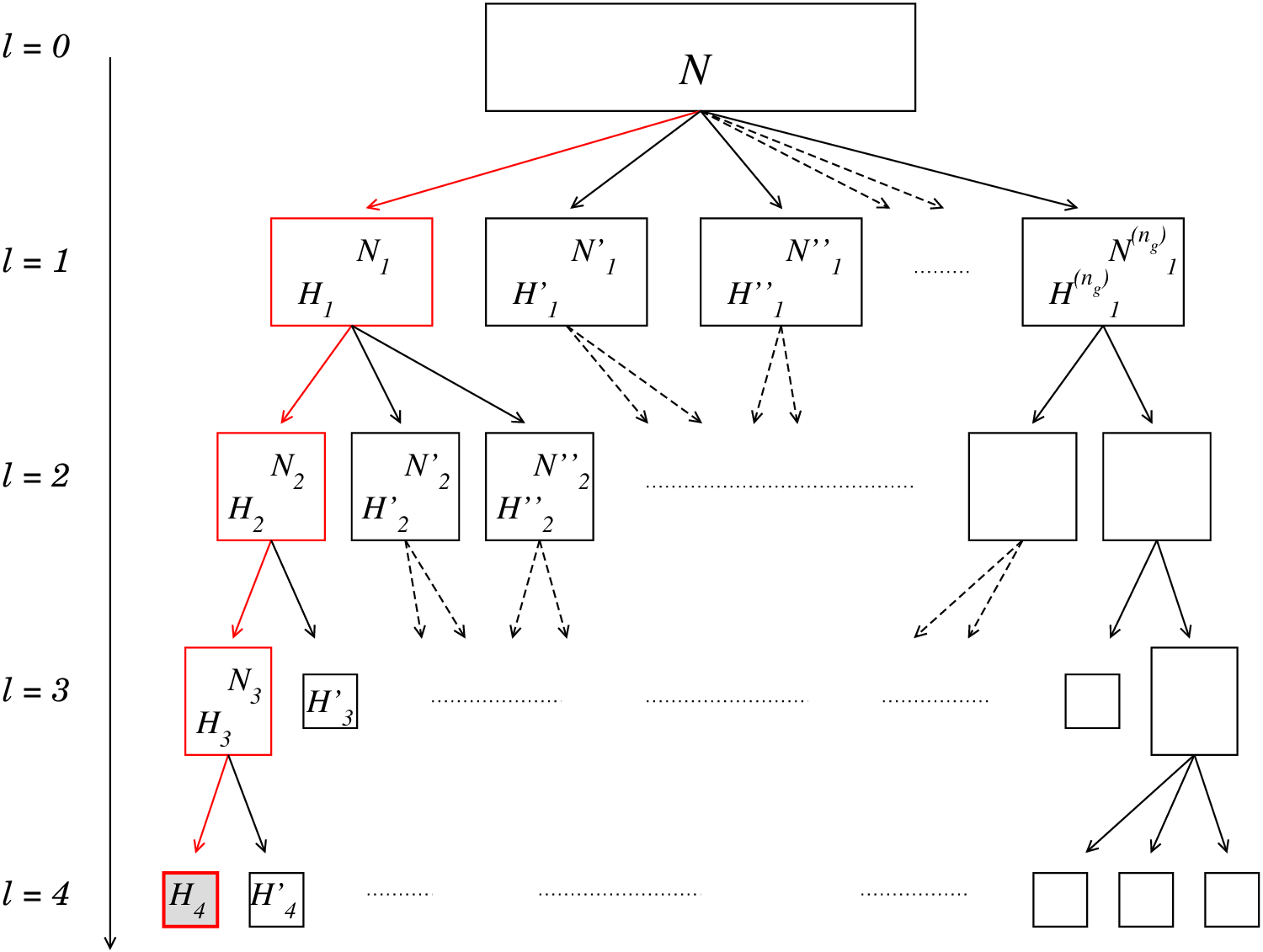}%{modele.eps}
\caption{\small (Color online) Phenomenological Model. At each level $\ell$, a group of size $N_l$ includes $\approx N_\ell^{1/3}$ subgroups of size $N_{\ell+1}$, i.e., $N_{\ell+1}=N_l^{2/3}\,(1+\eta)$ where $\eta$ represents a small noise.  The group-idiosyncrasy of the agent $i$ at its final level, $l_f(i)$, is simply called $H_{l_f}(i)=H(i)$ and $l_f(i)=d(i)$. For instance, the agent represented by the gray-filled box (ancestors marked in red) has the group-idiosyncrasy $H(i)=H_4$ and a distance $d(i)=4$.}
\label{fmodel}
\end{figure}

Due to the scale invariance of $\langle \tau \rangle$ and $\sigma^2_\tau$ we assume that $N$ individuals form $n_1$ groups (with $n_1\approx N^{1/3}$), of approximately $N_1$ individuals each, and each of these groups subdivide again in approximatively $n_2$ subgroups (with $n_2\approx N_1^{1/3}$) with $N_2$ individuals in each subgroup, and so on as is depicted in Fig.~\ref{fmodel}. Successive groups form a hierarchical organization, the constituency as a whole at the highest level. As the number of subgroups at each level fluctuates (see, for example, at level $l=1$, the group of $N_1$ individuals subdivides into $n_2=3$ subgroups but the neighboring group of $N'_1$ individuals subdivides into $n'_2=2$ subgroups), the length of all the branches (a branch stops when the group is too small, namely, one single individual) is not the same. We note $d(i)=l_f(i)$, where $l_f(i)$ is the final level of agent $i$, the total distance of an agent with respect to the highest level in the hierarchy, namely, the whole population $N$.

The intention to vote of the agent $i$, $C_i$, depends on three different components. First, the idiosyncratic part $H(i)$ reflects the intrinsic tendency of agents to vote or not, relative to their cultural, social, etc., background. We assume that agents belonging to the same group share the same idiosyncrasy. Furthermore, we assume that the groups at level $\ell +1$, issued from the same parent group at level $\ell$, share some common trait, while having some variability which accounts for the differences emerging at smaller scales. We thus write for the idiosyncrasy of a group at level $\ell +1$, $H_{\ell+1} = H_{\ell} + \gamma_{\ell+1}$, where $\gamma_\ell$ is random variable of zero mean and whose variance  decreases with  $\ell$. This decrease  stands for the fact that differences become smaller at a smaller scale.  The particular decrease law of  the variance is irrelevant. Here, for simplicity considerations, we have chosen a variance proportional to $1/2^{\ell}$.~\footnote{In this case, the variance of $H_{\ell}$ is proportional to $1/2+1/4+1/8+\dots+1/2^\ell$ which is $\approx1$ for $\ell\gg1$.}  We have tested different decreasing laws for the variance (proportional to $1/2^{2l}$, to $1/l$ or to $1/l^2$), and we have retrieved the same behavior.

Second, we assume that the agent $i$'s intention to vote also depends on his or her perceived link to the whole constituency. In other words, the intention to vote on a collective issue depends on how strong the feeling of belonging to that collectivity is. We represent this component by the final number of levels $d(i)$ that separates a citizen from the higher level, the whole constituency. Finally, the average intention to vote of the global population is represented by a global field $F_{nat}$ that is a constant over all the  municipalities of a country, for each election. Thus, the intention to vote of the agent $i$ finally reads
\begin{equation}
C_i = H(i) - d(i) + F_{nat}\,,
\label{econvic}
\end{equation}
and following Refs.~\cite{Bouchaud-rev,diffusive1,diffusive2}, the decision to vote (1) or not to vote (0) is given by $S_i = \Theta(C_i)$, where $\Theta$ is the Heaviside function.

%Fig. 11

The national average of the LTR does not play any role in the observed universal behavior of local elections. Thus, the average LTR  depends on the variance of the group-idiosyncratic term $H(i)$ and, if the distribution of $d(i)$ is not too broad (as it is the case here), on the average total distance of an agent with respect to the original population, $\langle d(i) \rangle (N)$. This distance is determined by the small randomness in the group distribution of each realization, so the only free parameter of the model is the variance of the distribution of idiosyncrasies at a given level. 

It can easily be seen that the mathematical expression of the LTR, $\tau$, obtained from the data, is particularly convenient for idiosyncrasies distributed according to a logistic distribution of zero mean. Let us recall that the probability that an agent, taken at random, will vote is $p=\frac{N_+}{N}$, for large $N$. Then, from the definition $\tau=\ln(\frac{N_+}{N-N_+})$, one gets: $p=\frac{1}{1+e^{-\tau}}$. On the other hand, let $H_{\ell}$, in Eq.~(\ref{econvic}), be a random variable distributed accordingly to a logistic distribution with zero mean and a scale parameter $s$, and let also, for simplicity, $d(i)=\overline{d}~ \forall i $. Then $p={\cal P}(C_i>0)=\frac{1}{1+e^{-(F-\overline{d})/s}}$ in the large $N$ limit. This yields $\tau=(F-\overline{d})/s$. For convenience, $\gamma$ can also be sampled from a logistic distribution without loss of generality. Then the only free parameter of the model  is  $s$, the scale parameter of the logistic distribution. The model agrees with the empirical observations for $s \approx 1$. Indeed, following the chosen form for the decrease with $\ell$ of the variance of $\gamma_\ell$, the scale parameter at the level $\ell\geq 1$ of the distribution of $\gamma_{\ell}$ is $s/\sqrt{2^\ell}$.\\

We have performed simulations over 200 realizations of the group distribution for a given $N$, each leading to a different realization of the hierarchy depicted in Fig.~\ref{fmodel}. For each realization and for each group at the level $\ell$, the size of subgroups at level $\ell+1$ is such that $N_{\ell+1}=N_\ell^{2/3}\,(1+\eta)$. The noise, $\eta$, is taken from a Gaussian distribution with zero mean, and standard deviation equal to $0.2$. This small irrelevant noise is introduced in order to avoid steplike curves. We have calculated, using Eq.~(\ref{econvic}) and the voting rule, the corresponding $\langle \tau \rangle$ and $\sigma^2_\tau$. The results, shown by the continuous line of Fig.~\ref{ftau}, have the same behavior as the  observed curves. Notice that the  model imposes that the simulated $\sigma^2_\tau (N)$ behaves as $1/N^\beta$ (with $\beta\approx0.3$) but interestingly, the obtained prefactor is very near to the one issued from observed data.

Several robustness tests have been performed. First, we have checked that these results are robust when varying the national field, $F_{nat}$, in the same range than the $\langle \tau \rangle (N)$ obtained from empirical curves (i.e., when the turnout rate spreads from $\approx 20\%$ up to $\approx 90\%$). Second, we have consider the scale parameter $s$ and we find that the results are in agreement with the empirical regularities for  $0.75\lesssim s \lesssim 1.15$ . Next, instead of directly introducing the noise in each subgroup (such that $N_{\ell+1}=N_\ell^{2/3}\,(1+\eta)$), we have directly injected the noise in the number of groups in order to account for the dispersion in the exponent observed in data. More precisely a group with $N_\ell$ agents at the level $\ell$ is directly split into $A\,N^\delta$ subgroups. Simulations retrieve empiric regularities when $0.28\lesssim \delta \lesssim 0.35$ with $A=1$, and $0.5\lesssim A \lesssim 2$ with $\delta=1/3$, which once again is consistent with the assumption $n_g \approx N^{1/3}$. Finally, we have also performed simulations replacing the logistic distribution of the group idiosyncrasy by a Gaussian or bimodal distribution, in order to verify that the results are independent from the particular choice of distribution.

%%%%%%%%%%%%%%%%%%%%%%%%%%%%%%%%%%%%%%%%%%%%%
\section{Conclusion}
We have found that the participation in a decision-making problem, when the $N$ agents have to decide about an issue that concerns only themselves as a group, shows a universal population-size effect. More specifically, we have shown that the average and variance logarithmic turnout rate (LTR) decrease in the same way with constituency size, irrespective of local particularities. This contrasts with the fact that the LTR exhibits no such regularities for national/European elections, when plotted as a function of the population of {\it the same set of constituencies}. This difference indicates that agents in each municipality may behave as if their group structure and cohesion becomes important when the object of the election concerns only themselves as a group. The observed behavior hints at a universal subdivision of the constituencies in $n_g \approx N^{\delta}$ subgroups with $\delta \approx 1/3$. We also find a similar universal behavior for the number of democratic representatives from municipal to national scales, as a function of the size $N$ of the constituency that has elected them. These results agree with the study of the candidature process~\cite{universality_candidates}.

A very simple phenomenological model, based on the existence of groups at all scales with the corresponding relation $n_g\approx N^{1/3}$ at each scale, quantitatively agrees with the observed behavior of the LTR. This idea of a hierarchy of group-idiosyncrasies has also been shown in Ref.~\cite{Luca_Valori}. Because in this work we calculate average values of the LTR over many (typically 100) municipalities of similar size, our results do not exclude that other factors might influence the participation rate of a given municipality as has been largely studied in Refs.~\cite{franklin,geys-empirical}.

This work calls for further interdisciplinary research as the origin of the regularities shown by the data remains unclear. Moreover, as these results seem to be related to the representation that $N$ agents have about themselves as a group, one could ask whether they reflect some intrinsic group dynamics that is common to the different  situations we analyzed, as has also been observed in anthropological studies~\cite{pop-size-oceania}, or, on the other hand, whether they reflect the way in which the agents (here humans) are used to dealing with group classification, as in taxonomic and soil classifications studied in Refs.~\cite{taxo_plants,taxo_soil}. The latter is suggested by the fact that, if in Eq.~(\ref{eq_repr}) we put $n_r=N^{1/3}$, the number of agents in one group, $N_1$, is the geometric mean of $N$ and $n_r$. This can be related to the logarithmic neural representation of numbers as discussed in Refs.~\cite{geometric,neuropsychological-constraints,sobkowicz-lognormal}.

\begin{acknowledgments}
CB thanks Brigitte Hazart --who decided to send  French {\it \'elections municipales} data, which triggered this study--, Eva Jakubcov\'a and Peter Kurz for help in gathering data; Yonathan Anson, Paul Chapron, Yves Couder, Stamatios Nicolis, Alexandra Reisigl-Borghesi, Giuilia Sandri, Lionel Tabourier and G\'erard Toulouse for enlightening discussions; and the {\it R\'egion IdF} for partially supporting this work through a DIM post-doctoral grant.
\end{acknowledgments}

%%%%%%%%%%%%%%%%%%%%%%%%%%%%%%%%%%%%%%
%%%%%%%%%%%%%%%%%%%%%%%%%%%%%%%%%%%%%%
%%%%%%%%%%%%%%%%%%%%%%%%%%%%%%%%%%%%%%
\renewcommand{\thesubsection}{A.\arabic{subsection}}
\setcounter{subsection}{0}  % reset counter 

%\appendix 
\section*{Appendix}
\subsection {Materials and methods: Data collection and treatment}%SI-A. 
\label{sdata}
Data used in this work can directly be downloaded from~\cite{download}.

%\paragraph*{Local elections study}
\subsubsection {Local elections study}
The studied countries are listed in Table~\ref{tdata}, and each election is referred to by its abbreviation and year. Data have been taken from Austria~\cite{data-mun-at}, Costa Rica~\cite{data-mun-cr}, the Czech Republic~\cite{data-mun-cz}, France~\cite{data-mun-fr}, Israel~\cite{data-mun-is}, Poland~\cite{data-mun-pl}, Portugal~\cite{data-mun-pt}, Romania~\cite{data-mun-ro}, Slovakia~\cite{data-mun-sk}, and Spain~\cite{data-mun-sp}.

When dealing with real data some compromise is sometimes needed in order to better fulfill the four conditions listed in the article. For instance, in France and Romania, where the local election may require a second round, only the first round is studied because it occurs in every municipality of the country. On the other hand, in Spain we have analyzed only data from the provinces (\textit{comunidades aut\'onomas}) that vote exclusively for local representatives on the election date: \textit{Andaluc\'{i}a}, \textit{Catalu\~na} and \textit{Galicia}. In Polish elections, the first round is excluded because it concerns local and regional elections and we have kept the second round which concerns only local elections. These considerations lead sometimes to a lower statistics, e.g. the polish second round takes place in $\approx$ 700 and 800 municipalities in the two analyzed elections. Finally the case of Israel is shown, despite its poor statistics, because it may help to understand the observed regularities, as discussed in the main text.

Some remarks on the notation used to describe the analyzed data:
\begin {itemize}
\item In Austria, though the election date may change from one province to another, we have labeled the elections in all the provinces with a single date.

\item Some Romanian electors, not registered in the \textit{lista electorala permanenta}, are able to vote. We do not consider those electors.

\item In one third of the municipalities of the Czech Republic, the local election takes place on the same day as Senate election, and in Romania, the municipal election and the election concerning representatives of {\it jude\c{t}e} (which are a constituted by a group of municipalities, except for large cities, where the corresponding {\it jude\c{t}} has the same size as the city) are simultaneous. Nevertheless we have integrated them in the general group (above the horizontal line of Table~\ref{tdata}), because we observe that this part of mixed data does not completely destroy the universal behavior of LTR.
\end{itemize}

Each point of Fig.~\ref{ftau} is obtained as the average over 100 municipalities of approximatively the same size, $N$, with the exception of France, Poland, and Israel, where each point represents the average over 200, 50, and 20 municipalities, respectively. Apart from the two extremal bins, (largest and smallest municipalities) each municipality is taken into account twice, i.e., belongs to two neighboring bins.

In order to get rid of electoral errors, we decide to exclude the extreme values of the LTR. Let us consider, for instance, some 100 municipalities of size $\approx N$ (as in Fig.~\ref{ftau}), and of LTR $\tau_i$ ($i=1,2,\dots,100$). We denote by $\mu_0$ and $\sigma_0$ the raw average and standard deviation of $\tau$ over these 100 municipalities. Then, the final average value $\langle \tau \rangle (N)$ and of the variance $\sigma^2_\tau (N)$ take only the municipalities $i$ such that $| \tau_i - \mu_0 | < 5 \sigma_0$ into account. It results that a very small fraction  of data is excluded (about $2/1000$).

%\clearpage
%\paragraph*{Study of the number of democratic representatives}
\subsubsection {Study of the number of democratic representatives}

We collect data of the number of representatives in democratic chambers at different scales, national, regional/state, and municipal, as a function of the total number of registered voters $N$ or, when unknown, the corresponding total population. 

For ``representative" we understand either a particular (elected) individual that represents a given group of people that has voted for him or her (as in the case where distinct constituencies exist and one particular representative talks in the name of each constituency), or the situation where a given number of representatives is attributed to the whole population. This approach distinguishes neither between councillors and deputies, nor between representatives directly elected from people constituency from those elected from their political parties.\\

\noindent
$\bullet$ {\bf Municipal Councils}\\
French local councillors ({\it conseillers municipaux}) in 2008, are taken from Ref.~\cite{repres-mun-fr}. The number of representatives of municipalities is directly given for Paris, Lyon, and Marseille, and for the remaining municipalities, it is given as a function of population range. The population of these towns comes from a 2011 national estimation~\cite{ign-2011}. A similar procedure is followed concerning data about municipal councillors of the seven {\it L\"ander}~\footnote{In Brandenburg and Schleswig-Holstein, we arbitrary put the population limit between {\it Gemeinden} and {\it St\"adten} at 50000.} in Germany where data were accessible~\cite{repres-mun-ge} and in Catalonia (in Spain). For the last Czech and Slovakian local elections, the total number of elected deputies/councillors for each municipality, as a function of the $N$ registered voters, is directly obtained from the corresponding official electoral data set.\\

\noindent
$\bullet$ {\bf Regional \footnote{Here by the word ``Regional" we mean an intermediate political organization between municipalities and the whole nation. These political units have different names and sizes in different countries, like ``states" in the United States, or {\it R\'egions} in France, {\it L\"ander} in Germany, etc.} Councils}\\
The analyzed cases are: nine Austrian Chambers (eight {\it Landtage} Parliaments and the Municipal Assembly of Vienna)~\cite{data-mun-at}, 27 Brazilian {\it Assembleias Legislativas Estaduais}, 13 Czech Regional Councils ({\it Zastupitelstva kraj\.{u}})\cite{repres-reg-cz}, 26 French {\it Parlements r\'egionaux}, 16 German {\it Landesparlamente}~\cite{repres-reg-ge}, 21 Italian {\it Consigli regionale}~\cite{repres-reg-it}, 16 Polish Voivodeship Sejmiks ({\it Sejmiki Wojew\'odzkie}), two {\it Assembleias Legislativas regionais} (for {\it A\c{c}ores} and {\it Madeira}) in Portugal~\cite{repres-reg-pt}, 41 Romanian County Councils ({\it Consilii Jude\c{t}ene}) and the Bucharest Municipal General Council~\cite{repres-reg-ro}, eight Regional Self-Governments in Slovakia~\cite{repres-reg-sk}, and 17 Spanish {\it Parlamentos auton\'omicos}. The number $N$ of registered voters is known in every region.\\

\noindent
$\bullet$ {\bf National chambers of Deputies}\\
Data concerning 181 democratic countries have been gathered using the sources in Refs.~\cite{repres-nat,repres-nat-africa}, where we have taken whenever possible, the information corresponding to the latest election. We do not consider National Constituent Assemblies. We take into account the total number $N$ of registered voters per country. In some few cases this number is estimated, using the same sources~\cite{repres-nat,repres-nat-africa}.\\

\noindent
$\bullet$ {\bf Evolution in time of the number of deputies in the U.S. House of Representatives}\\
The evolution in time of total number of deputies in the House of Representatives of the United States is the only empiric case where two important properties meet: (1) stability of the democratic regime over a long period, and (2) a significant change of the total population during this period (here the population has increased in two orders of magnitude from 1790 up to 2000). Data are taken from Refs.~\cite{histo-usa,histo-usa-2} or directly from Ref.~\cite{histo-usa-table}, where details on the considered population, etc., are also given.

Finally, general widespread data like the area and the nominal gross domestic product (GDP) per country used here, correspond to 2011/2012.

\subsection {A short historical investigation: ancient cities and the modern USA democracy }

We have investigated the number of ``representatives" in ancient cities which had an election procedure of election to choose their government, in order to check if they they also verified the behavior observed for present societies.

We show examples presenting the following characteristics: (1) they correspond to different historical periods, and (2) only assemblies or citizens who directly voted, acclaimed, etc., their ``representatives" are considered, in order to compare with our contemporary study. Unfortunately, most of the cases do not correspond to this criterion because either there were intermediate structures between the citizenry and those who held the legitimate authority (e.g., Geneva in the 17th century had the {\it Petit Conseil}, $\approx 25$ members, the {\it Conseil des deux-cents}, $\approx$ 200-300 members, and the {\it Conseil g\'en\'eral}, $\approx$ 1500 members) or the ``body of electors" was composed by members of different social status (e.g. the {\it popolo grasso} and the {\it popolo minuto} in Florence in the 14th century).

In Sparta, the citizens who corresponded to the {\it hoplites} elected five {\it ephors} for one year (\cite{sparte}, pp.~161-210). To have an approach as precise as possible of the number of Spartan citizens, we chose the 480~BC period, because it was the peak of the Sparta's military and political power, with 8000 {\it hoplites} according to Herodotus (\cite{herodote}, \cRM{7}, 234), and 371~BC when Sparta entered its decline with the defeat in Leuctra(~\cite{xenophon}, \cRM{6}, 4-15), where there were only 800 {\it hoplites} (\cite{sparte}, p.~269).

For the Athenian democracy, in the 5th century BC, we used the {\it Dictionary of Antiquity} published by the University of Oxford~\cite{dico-antiquite}. Thus, we could identify the population (\cite{dico-antiquite}, pp. 802-803) and with more specific entries about {\it ekklesia} (\cite{dico-antiquite}, p.~348) or the {\it boule} (\cite{dico-antiquite}, p.~154) we could clarify the overall organization of the Athens constitution~\cite{aristote-athenes}. On the {\it Pnyx} which was the official place where the citizens gathered, there could not be more than 6000 citizens because it was the maximum number of people who could really gather there to elect 10 {\it strategists} for one year.

Medieval Geneva had a population of about 5000 inhabitants in the 14th and 15th centuries~\cite{dico-suisse}, and all the male inhabitants could participate in the political life. Thus, about 2000 active citizens can be expected to have elected four {\it syndics} for one year (\cite{histoire-geneve}, pp.13-14).

As for the colony of Virginia, in 1619 there was established a {\it House of Burgesses} composed of 22 members directly elected by all the free men over 17, Protestants, and property owners~\cite{virginie1,virginie2}. As the population was 1277 people in 1624~\cite{wikipedia-virginie}, we can assume that the active citizens who participated in the elections of their representatives were about 600.

%Fig. 12
\begin{figure}[t]
\includegraphics[width=7.5cm, height=5.5cm, clip=true]{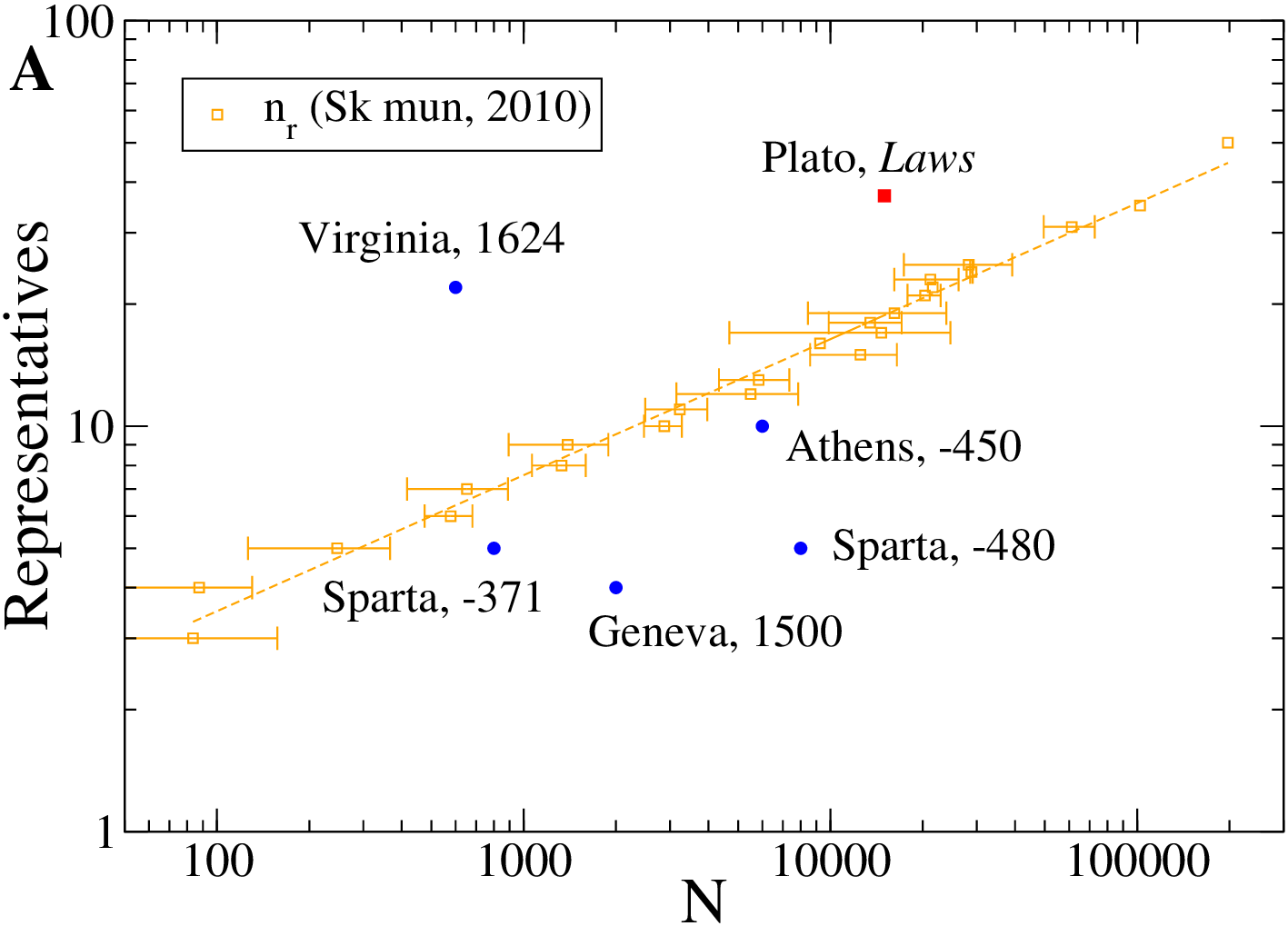}\\%{si-histoire.eps}
\includegraphics[width=7.5cm, height=5.5cm, clip=true]{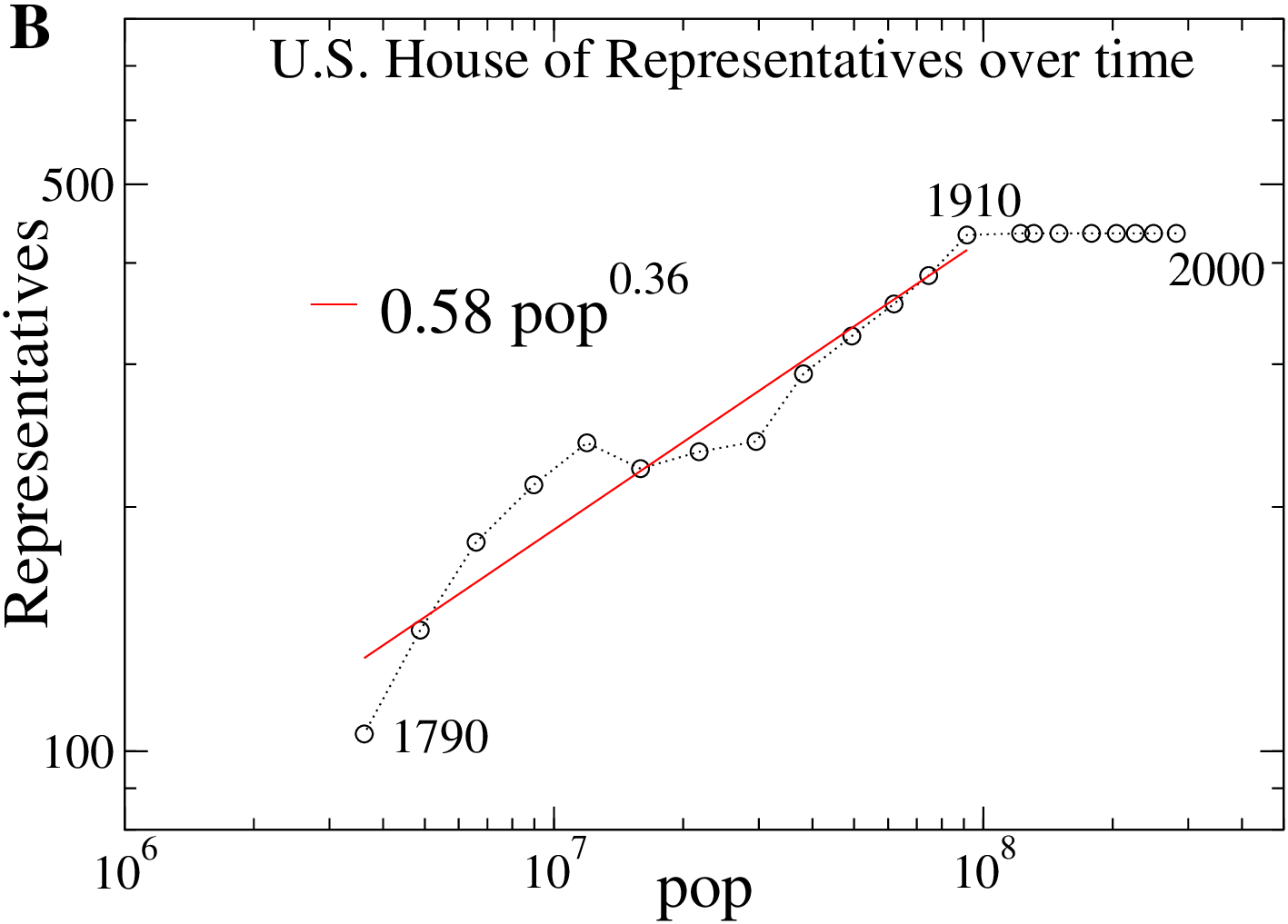}%{si-usa.eps}
\caption{\small (Color online) ({\bf A}) Total number of ``representatives'' with regard to the corresponding total number of citizens, $N$, from old cities and from a dialogue/utopia written in Antic Greece. Councillors from Slovakia Republic in current-day municipalities (in Fig.~\ref{frepres}A) are also plotted as a guide view. ({\bf B}) Evolution in time of $n_r$ with respect to the population in United States from 1790 to 2000 (each point corresponds to 10 years, except for 1920). The line corresponds to the power-law fit in the legend.}
\label{fsi-histoire}
\end{figure}

The last example comes from the dialogue/utopia {\it Laws} by Plato, and particularly in books \cRM{5} and \cRM{6}. The city of {\it Magnesia} should have 37 guardians of the laws from the around 15000 citizens (see Ref.~\cite{platon-lois}, section Annexes \cRM{1}, from which we have taken a low number of the estimated number of citizens).

Fig.~\ref{fsi-histoire}A plots the number of ``representatives" with respect to the number of citizens, $N$. It appears that regularities observed in present data (i.e. $n_r\approx N^\gamma$, with $\gamma\approx 1/3$) do not hold at all for these old cities. This is probably due to the fact that, though those ancient cities elected their ``representatives", the nature of their politic regime was different from our representative democracy.

On the other hand, a better fit is observed for the evolution of the number of congressmen in the United States with respect to the evolution of the population of the country over time (this country shows a large increase in population along with a stable democratic system). Fig.~\ref{fsi-histoire}B shows that it follows the same trend than regularities observed in present data until the beginning of the 20th century, when the number of congressmen last changed.


\begin{thebibliography}{99}
\bibitem{fortunato}C. Castellano, S. Fortunato, V. Loreto, Rev. Mod. Phys. {\bf 81}, 591-646 (2009).
\bibitem{Bouchaud-rev}J-.P. Bouchaud, J. Stat. Phys. {\bf 151}, 567-606 (2013).
\bibitem{costa_filho_scaling_vot}R. N. Costa Filho, M. P. Almeida, J. S. Andrade Jr., J. E. Moreira, Phys. Rev. E {\bf 60}, 1067-1068 (1999).
\bibitem{lyra_bresil_el}M. L. Lyra, U. M. S. Costa, R. N. Costa Filho, J. S. Andrade, Eur. Phys. Lett. {\bf 62}, 131-137 (2003).
\bibitem{fortunato2}S. Fortunato, C. Castellano, Phys. Rev. Lett. {\bf 99}, 138701 (2007).
\bibitem{daisy-model}H. Hern\'andez-Salda\~na, Physica A {\bf 388}, 2699-2704 (2009).
\bibitem{araripe_role_parties}L.E. Araripe, R.N. Costa Filho, Physica A {\bf 388}, 4167-4170 (2009).
\bibitem{diffusive1}C. Borghesi, J.-P. Bouchaud, Eur. Phys. J. B {\bf 75}, 395-404 (2010).
\bibitem{araujo_tactical_voting}N. A. M. Ara\'ujo, J. S. Andrade Jr., H. J. Herrmann, PLoS ONE {\bf 5}, e12446 (2010).
\bibitem{universality_candidates}M. C. Mantovani H. V. Ribeiro, M. V. Moro, S. Picoli Jr., R. S. Mendes, Eur. Phys. Lett. {\bf 96}, 48001 (2011).
\bibitem{diffusive2}C. Borghesi, J.-C. Raynal, J.-P. Bouchaud, PLoS ONE {\bf 7}, e36289 (2012).
\bibitem{weak-law}C. Borghesi, J. Chiche, J.-P. Nadal, PLoS ONE {\bf 7}, e39916 (2012).
\bibitem{fortunato-2012}A. Chatterjee, M. Mitrovi\'c, S. Fortunato, Sci. Rep. {\bf 3}, 1049 (2013); A. Chatterjee, M. Mitrovi\'c, S. Fortunato, Sci. Rep. {\bf 3}, 1155 (2013).
\bibitem{universality_candidates2}M. C. Mantovani, H. V. Ribeiro, E. K. Lenzi, S. Picoli, Jr. and R. S. Mendes, Phys. Rev. E {\bf 88}, 024802 (2013).% [5 pages]  %Engagement in the electoral processes: Scaling laws and the role of political positions
\bibitem{download}http://www.u-cergy.fr/fr/laboratoires/labo-lptm/donnees-de-recherche.html
\bibitem{these}C. Borghesi, \textit{Une \'etude de physique sur les \'elections} (Ed. Univ. Europ., Saarbr\"ucken, 2010).
\bibitem{klimek-irregularities}P. Klimek, Y. Yegorov, R. Hanel, S. Thurner, Proc. Natl. Acad. Sci. USA {\bf 109}, 16469-16473 (2012).
\bibitem{toulouse}G. Toulouse, J. Bok, Rev. Fran\c{c}. Socio. {\bf 19}, 391-406 (1978).
\bibitem{Luca_Valori}L. Valori, F. Picciolo, A. Allansdottir, D. Garlaschelli, {\it Proc. Natl. Acad. Sci. USA} {\bf 109}, 1068-1073 (2012).
\bibitem{franklin}N.M. Franklin, C. Van Der Eijk, D. Evans, M. Fotos, W. Hirczy De Mino, M. Marsh, B. Wessels {\it Voter Turnout and the Dynamics of Electoral Competition in Established Democracies since 1945} (Cambridge University Press, Cambridge, 2004).
\bibitem{geys-empirical}B. Geys, Electoral Studies {\bf 25}, 637-663 (2006).%{\it Explaining voter turnout: A review of aggregate-level research}.
\bibitem{pop-size-oceania}M.A. Kline, R. Boyd, Proc. R. Soc. London Ser. B {\bf 277}, 2559-2564 (2010).

\bibitem{taxo_plants}C. Caretta Cartozo, D. Garlaschelli, C. Ricotta, M. Barth\'el\'emy, G. Caldarelli, J. Phys. A {\bf 41}, 224012 (2008).
\bibitem{taxo_soil}J.J. Ibanez, R.W. Arnold, R.J. Ahrens, Ecol. Complex. {\bf 6}, 286-293 (2009).
\bibitem{geometric}S. Dehaene, V. Izard, E. Spelke, P. Pica, Science {\bf 320}, 1217-1220 (2008).%{\it Log or Linear? Distinct Intuitions of the Number Scale in Western and Amazonian Indigene Cultures}. 
\bibitem{neuropsychological-constraints}C. Gros, G. Kaczor, D. Markovi\'c, Eur. Phys. J. B {\bf 85}:28 (2012).%{\it Neuropsychological constraints to human data production on a global scale}
\bibitem{sobkowicz-lognormal}P. Sobkowicz, M. Thelwall, K. Buckley, G. Paltoglou, A.Sobkowicz, EPJ Data Sci. {\bf 2}:2 (2013).%{it Lognormal distributions of user post lengths in Internet discussions - a consequence of the Weber-Fechner law?}

%%%%%%%%%%%%%%%%%%%%%%%%%%%%%%%%
%%%%%%%%%%%%%%%%%%%%%%%%%%%%%%%%
%\bibitem{download_s}http://www.u-cergy.fr/fr/laboratoires/labo-lptm/donnees-de-recherche.html
%for local_elections data
\bibitem{data-mun-at}http://www.bmi.gv.at/cms/BMI\_wahlen/
\bibitem{data-mun-cr}http://www.consulta.tse.go.cr/estadisticas\_elecciones.htm
\bibitem{data-mun-cz}http://www.volby.cz %http://www.czso.cz
\bibitem{data-mun-fr}http://www.interieur.gouv.fr/Elections/Les-resultats
\bibitem{data-mun-is}http://www.moin.gov.il/Subjects/Bchirot/Pages\linebreak/resultmekomi.aspx
\bibitem{data-mun-pl}http://pkw.gov.pl/wyniki-wyborow-i-referendow/wybory-i-referenda.html
\bibitem{data-mun-pt}http://eleicoes.cne.pt/
\bibitem{data-mun-ro}http://www.beclocale2012.ro/rezultate.html
\bibitem{data-mun-sk}http://portal.statistics.sk/showdoc.do?docid=5673
\bibitem{data-mun-sp}http://www.infoelectoral.mir.es/min/
%for SI-representants
\bibitem{repres-mun-fr}http://www.politiquemania.com/nombre-conseillers-municipaux-par-commune.html
\bibitem{ign-2011}http://professionnels.ign.fr/rgc%Institut G\'eographique National, {\it R\'epertoire g\'eographique des communes}\\
\bibitem{repres-mun-ge}http://www.wahlrecht.de/kommunal/
\bibitem{repres-reg-cz}http://www.volby.cz/index\_en.htm
\bibitem{repres-reg-ge}http://www.wahlrecht.de/ergebnisse/index.htm
\bibitem{repres-reg-it}http://elezionistorico.interno.it/ %and http://www.repubblica.it/static/speciale/2010/elezioni/regionali/
\bibitem{repres-reg-pt}http://www.cne.pt/
\bibitem{repres-reg-ro}http://www.beclocale2012.ro/
\bibitem{repres-reg-sk}http://portal.statistics.sk/showdoc.do?docid=3090
\bibitem{repres-nat}http://www.ipu.org/parline/
\bibitem{repres-nat-africa}http://africanelections.tripod.com/
\bibitem{histo-usa}http://www.thirty-thousand.org/pages/analyses.htm
\bibitem{histo-usa-2}https://pantherfile.uwm.edu/margo/www/apport\linebreak/datasets.htm
\bibitem{histo-usa-table}www.uwm.edu/~margo/apport/apportionment1.pdf
%for SI-histoire
\bibitem{sparte}Edmond Levy, {\it Sparte, Histoire politique et sociale jusqu'\`a la conqu\^ete romaine} (Ed. Seuil, Paris, 2003).
\bibitem{herodote}H\'erodote, {\it L'enqu\^ete} (translated by A. Barguet, Ed. Gallimard, Paris, 1990).
\bibitem{xenophon}X\'enophon, {\it Hell\'eniques} (translated by E. Talbot, Ed. Gallimard, Paris, 2012).
\bibitem{dico-antiquite}M. C. Howatson, University of Oxford, {\it Dictionnaire de l’Antiquit\'e} (Collection Bouquin, Paris, 1993).
\bibitem{aristote-athenes}Aristote, \textit{Constitution d'Ath\`enes} (translated by G. Mathieu, B. Haussoulier, and C. Moss\'e, Ed. Les Belles-Lettres, Paris, 1996).
\bibitem{dico-suisse}{\it Dictionnaire historique de la Suisse}, www.hls-dhs-dss.ch/textes/f/F2903.php
\bibitem{histoire-geneve}L. Binz, {\it Br\`eve histoire de Gen\`eve} (Chancellerie de Gen\`eve, Geneva, 2000).
\bibitem{virginie1}http://www.49online.org/webpages/nschumacher\linebreak/index.cfm?subpage=506084
\bibitem{virginie2}http://www.statemaster.com/encyclopedia/House-of-Burgesses
\bibitem{wikipedia-virginie}http://en.wikipedia.org/wiki/History\_of\_Virginia
\bibitem{platon-lois}Platon, {\it Les Lois} (translated by L. Brisson and J.-F. Pradeau, Ed. Flammarion, Paris, 2006).
\end{thebibliography}
\end{document}